\documentstyle[prd,aps]{revtex}

\newcommand{\bea}{\begin{eqnarray}}
\newcommand{\eea}{\end{eqnarray}}

\begin{document}

\draft

\title{Cosmological nonlinear hydrodynamics with post-Newtonian corrections}
\author{Jai-chan Hwang${}^{(a)}$, Hyerim Noh${}^{(b)}$,
        and Dirk Puetzfeld${}^{(c)}$}
\address{${}^{(a)}$ Department of Astronomy and Atmospheric Sciences,
                    Kyungpook National University, Taegu, Korea \\
         ${}^{(b)}$ Korea Astronomy and Space Science Institute,
                    Daejon, Korea \\
         ${}^{(c)}$ Department of Physics and Astronomy,
                    Iowa State University, Ames, IA, 50011, USA
         }
\maketitle

\begin{abstract}

The purpose of this paper is to present general relativistic
cosmological hydrodynamic equations in Newtonian-like forms using
the post-Newtonian (PN) method. The PN approximation, based on the
assumptions of weak gravitational fields and slow motions, provides
a way to estimate general relativistic effects in the fully
nonlinear evolution stage of the large-scale cosmic structures. We
extend Chandrasekhar's first order PN (1PN) hydrodynamics based on
the Minkowski background to the Robertson-Walker background. We {\it
assume} the presence of Friedmann's cosmological spacetime as a
background. In the background we include the three-space curvature,
the cosmological constant and general pressure; we show that our 1PN
approach is successful only for spatially flat cosmological
background. In the Newtonian order and 1PN order we include general
pressure, stress, and flux. The Newtonian hydrodynamic equations
appear naturally in the 0PN order. The spatial gauge degree of
freedom is fixed in a unique manner and the basic equations are
arranged without taking the temporal gauge condition. In this way we
can conveniently try alternative temporal gauge conditions. We
investigate a number of temporal gauge conditions under which all
the remaining variables are equivalently gauge-invariant. Our aim is
to present the fully nonlinear 1PN equations in a form suitable for
implementation in conventional Newtonian hydrodynamic simulations
with minimal extensions. The 1PN terms can be considered as
relativistic corrections added to the well known Newtonian
equations. The proper arrangement of the variables and equations in
combination with suitable gauge conditions would allow the possible
future 1PN cosmological simulations to become more tractable. Our
equations and gauges are arranged for that purpose. We suggest ways
of controlling the numerical accuracy. The typical 1PN order terms
are about $10^{-6} \sim 10^{-4}$ times smaller than the Newtonian
terms. However, we cannot rule out possible presence of cumulative
effects due to the time-delayed propagation of the relativistic
gravitational field with finite speed, in contrast to the Newtonian
case where changes in the gravitational field are felt
instantaneously. The quantitative estimation of such effects is left
for future numerical simulations.

\end{abstract}


\tableofcontents

%
%
\section{Introduction}
                                           \label{sec:Introduction}

The nonlinear evolution of large-scale cosmic structures is usually
investigated within the framework of Newtonian theory either in
analytic studies or numerical simulations. Without investigating
general relativistic effects, however, it is not clear whether the
Newtonian theory is sufficient to handle such large-scale
cosmological structures. If we regard Einstein's gravity theory to
be the correct framework on such cosmic scales, the relativistic
effects should exist always. The point is to which level we can
practically ignore relativistic correction terms considering
currently available levels of observations and experiments. In case
relativistic or nonlinear effects are not large, we have two widely
known ways to estimate relativistic effects: one is the perturbation
approach \cite{Lifshitz-1946,Linear-perturbations,Bardeen-1980}, and
the other one is the post-Newtonian (PN) approximation
\cite{Fock-1964,Chandrasekhar-1965,Chandrasekhar-1969,Chandrasekhar-Nutku-1969,Chandrasekhar-Esposito-1970,Damour-1987}.

In the perturbation study, we recently have investigated the weakly
nonlinear regimes based on Einstein's gravity. We showed that,
except for the gravitational wave coupling, the relativistic
scalar-type perturbations coincide exactly with the Newtonian ones
up to the second order \cite{HN-second}. The pure relativistic
correction terms appear in the third order. These terms turn out to
be independent of the horizon scale and small ($\sim 10^{-5}$ order)
compared with the second-order Newtoninan/relativistic terms
\cite{HN-third}. Our studies have shown, from the view point of
Einstein's gravity, that Newtonian gravity is practically reliable
near the horizon scale where structures are supposed to be weakly
nonlinear. Therefore, justifying its use in current large-scale
numerical simulations which cover the Hubble volume. Considering the
action-at-a-distance nature of Newton's gravity our result is a
rather surprising one because the relativistic perturbation theory
is applicable on {\it all scales} including the super-horizon scale.
Thus, our perturbation study ensures that to the weakly nonlinear
order Newtonian theory is reliable even near and beyond the horizon
scale.

Now, how about situation in the nonlinear regime on scales smaller
than the current horizon? On such scales the structures could be in
a fully nonlinear stage. However, if the relativistic gravity effect
is small we can apply another approximation scheme which is well
developed to handle isolated bodies in Einstein's theory: the PN
approximation. The PN approximation has been important to test
Einstein's gravity. It presents the relativistic equations in a form
similar to the well known Newtonian equations. In the PN method, by
assuming that the relativistic effects are small, we expand
relativistic corrections in powers of $v/c$. In nearly virialized
systems we have $GM/(Rc^2) \sim (v/c)^2$. Thus, the PN approximation
is applicable in the slow motion and weak gravitational field
regime. The first order PN (1PN) approximation corresponds to adding
relativistic correction terms of the order $(v/c)^2$ to the
Newtonian order terms. In this approach we recover the well known
Newtonian hydrodynamics equations in the 0PN order
\cite{Chandrasekhar-1965}. The PN approximation is suitable to study
systems in which Newtonian gravity has a dominant role, while the
relativistic effects are small but non-negligible. Well known
applications include the precession of Mercury's perihelion, and
various solar system tests of Einstein's gravity theory like the
light deflection \cite{Will-1981}. Recent successful applications
include the generation of gravitational waves from compact binary
objects, and the weakly relativistic evolution stages of isolated
systems of celestial bodies \cite{Asada-Futamase-1997}. Another
important application is relativistic celestial mechanics which is
required by recent technology driven precise measurements of the
solar system bodies \cite{Brumberg-1991,Soffel-2003}. Notice that
all these applications are based on the PN approximation of isolated
systems assuming the Minkowski background spacetime.

In this work on the cosmological PN approximation we assume the
presence of a Robertson-Walker background. Thus, we have the
Friedmann world model built in as the background spacetime. We will
show that the Newtonian hydrodynamic equations come out naturally in
the 0PN order which is the Newtonian limit. In the context of
large-scale cosmic structures PN effects could be essentially
important in handling gravitational waves and gravitational lensing
effects. It is well known that Newtonian order treatment of the
gravitational lensing shows only half of the result from Einstein's
theory which is due to ignoring the 1PN correction in the metric. We
also anticipate that PN correction terms could affect the dynamic
evolution of large-scale structures, especially considering the
action-at-a-distance nature of the Newtonian theory. This can be
compared with the relativistic situation in which the gravitational
field propagates with finite speed. In this work we present the
complete set of equations to study 1PN effects in a cosmological
context. The PN approach can be compared with our previous studies
of the weakly nonlinear regime based on the perturbation approach
which is applicable in the fully relativistic regime and on all
scales. In comparison, although the PN equations are applicable in
weak gravity regions inside the horizon, these are applicable in
fully nonlinear situation. Thus, the two approaches are
complementary in enhancing our understanding of the relativistic
evolutionary aspects of the large-scale structure in the universe.

In the PN approximation we attempt to study the equations of motion
and the field equations in Einstein's gravity in a Newtonian way as
closely as possible. The Newtonian and post-Newtonian equations of
motion follow from the energy-momentum conservation. The Newtonian
order potential and 1PN order metric variables are determined by
Einstein's equations in terms of the Newtonian matter and potential
variables. Thus, the metric (or relativistic) contributions are
reinterpreted as small correction terms to the well known Newtonian
hydrodynamic equations. The Newtonian equations naturally follow in
the 0PN order. The form of 1PN equations is affected by our choice
of the gauge conditions. In this work we take unique spatial gauge
conditions which fix the spatial gauge modes completely; under these
spatial gauge conditions the remaining variables are all
equivalently spatially gauge-invariant, see Sec.\ \ref{sec:Gauges}.
The temporal gauge condition (slicing condition) will be deployed to
handle the mathematical treatment of the equations conveniently. We
show how to choose the temporal gauge condition which also removes
the temporal gauge mode completely. Under each of such temporal
gauge conditions the remaining variables are equivalently (spatially
and temporally) gauge-invariant. We arrange the equations and gauges
so that the final form of equations is suitable for numerical
implementation in conventional cosmological hydrodynamic
simulations.

We present basic quantities and the derivation of our 1PN equations
in Sec.\ \ref{sec:Quantities} and \ref{sec:Derivations}. If the
reader is more interested in the application of the 1PN equations,
the basic set of cosmological 1PN equations is summarized in Sec.\
\ref{sec:Equations}. The gauge issue is expounded in Sec.\
\ref{sec:Gauges}. We discuss our results and several unresolved
issues in Sec.\ \ref{sec:Discussion}.

%
%
\section{Basic quantities}
                                           \label{sec:Quantities}

\subsection{Curvature}

As the metric we take \bea
   & & \tilde g_{00} \equiv - \left[ 1 - {1 \over c^2} 2 U
       + {1 \over c^4} \left( 2 U^2 - 4 \Phi \right) \right]
       + {\cal O}^{-6},
   \nonumber \\
   & & \tilde g_{0i} \equiv - {1 \over c^3} a P_i + {\cal O}^{-5},
   \nonumber \\
   & & \tilde g_{ij}
       \equiv a^2 \left( 1 + {1 \over c^2} 2 V \right) \gamma_{ij}
       + {\cal O}^{-4},
   \label{metric}
\eea where $x^0 \equiv c t$, and $a(t)$ is the cosmic scale factor
of the background Friedmann world model. Indices $a, b, c, \dots$
indicate spacetime, and $i,j,k, \dots$ indicate space. Tildes
indicate spacetime covariant quantities, i.e., spacetime indices of
quantities with tilde are raised and lowered with the spacetime
metric $\tilde g_{ab}$. The spatial index of $P_i$ is based on
$\gamma_{ij}$ in raising and lowering indices with $\gamma^{ij}$, an
inverse of $\gamma_{ij}$. $\gamma_{ij}$ is the comoving
(time-independent) spatial part of the Robertson-Walker metric, see
Eq.\ (2) in \cite{HN-GGT-2005}. In a flat Robertson-Walker
background $\gamma_{ij}$ becomes $\delta_{ij}$ if we take Cartesian
coordinates. Compared with our previous notations used in
perturbation studies in \cite{NL}, we set the comoving part of the
three-space background metric $g^{(3)}_{ij} \equiv \gamma_{ij}$ and
use part of the Latin indices to indicate the space. We are
following Chandrasekhar and Nutku's notation
\cite{Chandrasekhar-1965,Chandrasekhar-Nutku-1969} extended to the
cosmological situation, see also Fock \cite{Fock-1964}. The $2U/c^2$
term in $\tilde g_{00}$ gives the Newtonian limit, and if we ignore
all the Newtonian and post-Newtonian correction terms we have the
Robertson-Walker spacetime. Thus, our PN formulation is built on the
cosmological background spacetime. ${\cal O}^{-n}$ indicates
$(v/c)^{-n}$ and higher order terms that we ignore. The expansion in
Eq.\ (\ref{metric}) is valid to 1PN order \cite{Chandrasekhar-1965}.
Dimensions are as follows \bea
   & & [\tilde g_{ab}] = [\tilde g^{ab}] = 1, \quad
       [\gamma_{ij}] = [\gamma^{ij}] = 1, \quad
       [a] = 1, \quad
       [c] = L{\cal T}^{-1}, \quad
       [U] = [V] = [c^2], \quad
       [P_i] = [P^i] = [c^3], \quad
       [\Phi]= [c^4],
\eea where $L$ and ${\cal T}$ indicate the length and the time
dimensions, respectively.

In our metric convention in Eq.\ (\ref{metric}) we have ignored the
possible presence of ${1 \over c^2} \left( 2 C_{,i|j} + C_{i|j} +
C_{j|i} \right)$ like terms in $\tilde g_{ij}$ with $C^i_{\;|i}
\equiv 0$ by choosing the spatial $C$-gauge conditions ($C \equiv 0
\equiv C_i$) which remove the spatial gauge mode completely; this
will be explained in Sec.\ \ref{sec:Gauges}; a vertical bar
indicates the covariant derivative based on $\gamma_{ij}$. Under
such spatial gauge conditions, we can regard all our remaining PN
variables as equivalently spatially gauge-invariant ones, see Sec.\
\ref{sec:Gauges}. We still have a freedom to take the temporal gauge
condition which will be chosen later depending on the mathematical
simplification or feasibility of physical interpretation of the
problem under consideration, see Sec.\ \ref{sec:Gauges}: in the
perturbation approach we termed this a gauge-ready approach
\cite{Hwang-1991}. We also have ignored the spatially
tracefree-transverse part of the metric ($C_{ij}$ with $C^i_i \equiv
0 \equiv C^j_{i|j}$) because gravitational waves are known to show
up only from the 2.5PN order \cite{Chandrasekhar-Esposito-1970}.

The inverse metric becomes \bea
   & & \tilde g^{00} = - \left[ 1 + { 1\over c^2} 2 U
       + {1 \over c^4} \left( 2 U^2 + 4 \Phi \right) \right]
       + {\cal O}^{-6},
   \nonumber \\
   & & \tilde g^{0i} = - {1 \over c^3} {1 \over a} P^i + {\cal O}^{-5},
   \nonumber \\
   & & \tilde g^{ij} = {1 \over a^2} \left( 1 - {1 \over c^2} 2 V \right)
       \gamma^{ij}
       + {\cal O}^{-4}.
\eea The determinant of the metric tensor $\tilde g$ is \bea
   & & \sqrt{ - \tilde g} = a^3 \sqrt{\gamma}
       \left[ 1 + {1 \over c^2} \left( 3 V - U \right)
       + {\cal O}^{-4} \right],
\eea where $\gamma$ is the determinant of $\gamma_{ij}$.

The connection is \bea
   \tilde \Gamma^0_{00}
   &=& - {1 \over c^3} \dot U
       + {1 \over c^5} \left( - 2 \dot \Phi
       + {1 \over a} P^i U_{,i} \right)
       + L^{-1} {\cal O}^{-7},
   \nonumber \\
   \tilde \Gamma^0_{0i}
   &=& - {1 \over c^2} U_{,i}
       - {1 \over c^4} \left( 2 \Phi_{,i}
       + \dot a P_i \right)
       + L^{-1} {\cal O}^{-6},
   \nonumber \\
   \tilde \Gamma^0_{ij}
   &=& a^2 \left\{ {1 \over c} {\dot a \over a} \gamma_{ij}
       + {1 \over c^3} \left[ \left( \dot V
       + 2 {\dot a \over a} \left( U + V \right) \right) \gamma_{ij}
       + {1 \over a} P_{(i|j)} \right] \right\}
       + L^{-1} {\cal O}^{-5},
   \nonumber \\
   \tilde \Gamma^i_{00}
   &=& {1 \over a^2} \left\{ - {1 \over c^2} U^{,i}
       + {1 \over c^4} \left[
       2 \left( U + V \right) U^{,i} - 2 \Phi^{,i}
       - \left( a P^i \right)^\cdot
       \right] \right\}
       + L^{-1} {\cal O}^{-6},
   \nonumber \\
   \tilde \Gamma^i_{0j}
   &=& {1 \over c} {\dot a \over a} \delta^i_j
       + {1 \over c^3} \left[ \dot V \delta^i_j
       - {1 \over 2 a} \left( P^i_{\;\; |j} - P_j^{\;\; |i} \right) \right]
       + L^{-1} {\cal O}^{-5},
   \nonumber \\
   \tilde \Gamma^i_{jk}
   &=& \Gamma^{(\gamma)i}_{\;\;\;\;\; jk}
       + {1 \over c^2} \left( V_{,k} \delta^i_j
       + V_{,j} \delta^i_k - V^{,i} \gamma_{jk} \right)
       + L^{-1} {\cal O}^{-4},
\eea where $\Gamma^{(\gamma)i}_{\;\;\;\;\; jk}$ is the connection
based on $\gamma_{ij}$; we introduce $P_{(i|j)} \equiv {1 \over 2}
\left( P_{i|j} + P_{j|i} \right)$ and $P_{[i|j]} \equiv {1 \over 2}
\left( P_{i|j} - P_{j|i} \right)$. We have $U_{,0} = {1 \over c}
{\partial U \over
\partial t} \equiv {1 \over c} \dot U$.

The Riemann curvature is \bea
   \tilde R^0_{\;\; 00i}
   &=& - {1 \over c^5} \left( \ddot a P_i
       + {1 \over a} U_{,i|j} P^j \right)
       + L^{-2} {\cal O}^{-7},
   \nonumber \\
   \tilde R^0_{\;\; 0ij}
   &=& L^{-2} {\cal O}^{-6},
   \nonumber \\
   \tilde R^0_{\;\; i0j}
   &=& {1 \over c^2} \left( a \ddot a \gamma_{ij}
       + U_{,i|j} \right)
       + {1 \over c^4} \Bigg\{
       a^2 \left[ \dot V + 2 {\dot a \over a} \left( U + V \right)
       \right]^\cdot \gamma_{ij}
       + \left[ - a \dot a \dot U
       + 2 \dot a^2 \left( U + V \right) \right] \gamma_{ij}
   \nonumber \\
   & &
       - 2 U_{,(i} V_{,j)}
       - U_{,i} U_{,j}
       + U^{,k} V_{,k} \gamma_{ij}
       + 2 \Phi_{,i|j}
       + \left( a P_{(i|j)} \right)^\cdot
       \Bigg\}
       + L^{-2} {\cal O}^{-6},
   \nonumber \\
   \tilde R^0_{\;\; ijk}
   &=& {1 \over c^3} a^2 \left[
       2 \left( \dot V + {\dot a \over a} U \right)_{,[j}
       \gamma_{k]i}
       - {1 \over a} P_{[j|k]i}
       + {2 K \over a} \gamma_{i[j} P_{k]} \right]
       + L^{-2} {\cal O}^{-5},
   \nonumber \\
   \tilde R^i_{\;\; 00j}
   &=& {1 \over c^2} \left( {\ddot a \over a} \delta^i_j
       + {1 \over a^2} U^{,i}_{\;\;\;|j} \right)
       + {1 \over c^4} \Bigg\{
       \left[ \ddot V + {\dot a \over a} \left( \dot U + 2 \dot V
       \right) \right] \delta^i_j
       - {1 \over a^2} \Big[ 2 \left( U + V \right)
       U^{,i}_{\;\;\;|j}
   \nonumber \\
   & &
       + U^{,i} U_{,j}
       + U^{,i} V_{,j}
       + U_{,j} V^{,i}
       - U^{,k} V_{,k} \delta^i_j
       - 2 \Phi^{,i}_{\;\;\;|j} \Big]
       + {1 \over 2 a^2} \left[ a \left(
       P^i_{\;\; |j} + P_j^{\;\;|i} \right) \right]^\cdot
       \Bigg\}
       + L^{-2} {\cal O}^{-6},
   \nonumber \\
   \tilde R^i_{\;\; 0jk}
   &=& {1 \over c^3} \left[ 2 \left( \dot V
       + {\dot a \over a} U \right)_{,[j} \delta^i_{k]}
       - {1 \over a} P_{[j|k]}^{\;\;\;\;\;\; i} \right]
       + L^{-2} {\cal O}^{-5},
   \nonumber \\
   \tilde R^i_{\;\; j0k}
   &=& {1 \over c^3} \left[ \dot V_{,j} \delta^i_k
       - \dot V^{,i} \gamma_{jk}
       + {\dot a \over a} \left( U_{,j} \delta^i_k
       - U^{,i} \gamma_{jk} \right)
       + {1 \over 2 a} \left( P^i_{\;\; |j}
       - P_j^{\;\; |i} \right)_{|k} \right]
       + L^{-2} {\cal O}^{-5},
   \nonumber \\
   \tilde R^i_{\;\; jkl}
   &=& - 2 K \gamma_{j[k} \delta^i_{l]}
       + {1 \over c^2} 2 \left( - \dot a^2 \gamma_{j[k}
       \delta^i_{l]}
       + V_{,j|[k} \delta^i_{l]}
       - V^{,i}_{\;\;\; |[k} \gamma_{l]j} \right)
       + L^{-2} {\cal O}^{-4}.
\eea It is convenient to have \bea
   & & P^i_{\;\;|jk} \equiv P^i_{\;\;|kj}
       - R^{(\gamma)i}_{\;\;\;\;\;\;\; ljk} P^l, \quad
       P_{i|jk} = P_{i|kj}
       + R^{(\gamma)l}_{\;\;\;\;\;\;\; ijk} P_l,
   \nonumber \\
   & &
       R^{(\gamma)i}_{\;\;\;\;\;\;\; jkl}
       = K \left( \delta^i_k \gamma_{jl} - \delta^i_l \gamma_{jk}
       \right), \quad
       R^{(\gamma)}_{ij} = 2 K \gamma_{ij}, \quad
       R^{(\gamma)} = 6 K,
\eea where $K$ indicates the comoving (time-independent) part of
background spatial curvature with dimension $[K] = L^{-2}$. The
Ricci curvature and the scalar curvature become \bea
   \tilde R^0_0
   &=& {1 \over c^2} \left( 3 {\ddot a \over a}
       + {\Delta \over a^2} U \right)
       + {1 \over c^4} \left\{
       3 \ddot V
       + 3 {\dot a \over a} \left( \dot U + 2 \dot V \right)
       + 6 {\ddot a \over a} U
       - {1 \over a^2} \left[ U^{,i} \left( U - V \right)_{,i}
       + 2 V \Delta U
       - 2 \Delta \Phi
       - \left( a P^i_{\;\;|i} \right)^\cdot \right] \right\}
   \nonumber \\
   & &
       + L^{-2} {\cal O}^{-6},
   \nonumber \\
   \tilde R^0_i
   &=& {1 \over c^3} \left[ 2 \left( \dot V + {\dot a \over a} U
       \right)_{,i}
       + {1 \over 2 a} \left( P^j_{\;\;|ji} - \Delta P_i
       - 2 K P_i \right)
       \right]
       + L^{-2} {\cal O}^{-5},
   \nonumber \\
   \tilde R^i_0
   &=& - {1 \over c^3} {1 \over a^2} \left[
       2 \left( \dot V + {\dot a \over a} U \right)^{,i}
       + {1 \over 2 a} \left( P^{j \;\;\; i}_{\;\;|j}
       - \Delta P^i + 2 K P^i \right) \right]
       + L^{-2} {\cal O}^{-5},
   \nonumber \\
   \tilde R^i_j
   &=& {2 K \over a^2} \delta^i_j
       + {1 \over c^2} \left[
       \left( {\ddot a \over a} + 2 {\dot a^2 \over a^2}
       - {\Delta + 4 K \over a^2} V \right) \delta^i_j
       + {1 \over a^2} \left( U - V \right)^{|i}_{\;\;\; j} \right]
       + L^{-2} {\cal O}^{-4},
   \nonumber \\
   \tilde R
   &=& {6 K \over a^2}
       + {1 \over c^2} \left[ 6 \left( {\ddot a \over a}
       + {\dot a^2 \over a^2} \right)
       + 2 {\Delta \over a^2} U
       - 4 {\Delta + 3 K \over a^2} V \right]
       + L^{-2} {\cal O}^{-4},
   \label{Rab}
\eea where $\Delta$ is a Laplacian operator based on $\gamma_{ij}$.

The Weyl curvature is introduced as \bea
   & & \tilde C^a_{\;\; bcd} \equiv
        \tilde R^a_{\;\; bcd}
        - {1 \over 2} \left( \delta^a_c \tilde R_{bd}
        - \delta^a_d \tilde R_{bc}
        + \tilde g_{bd} \tilde R^a_c
        - \tilde g_{bc} \tilde R^a_d \right)
        + {1 \over 6} \tilde R \left( \delta^a_c \tilde g_{bd}
        - \delta^a_d \tilde g_{bc} \right).
\eea To 1PN order we have \bea
   \tilde C^0_{\;\; 00i}
   &=& - {1 \over c^3} {K \over a} P_i
       + L^{-2} {\cal O}^{-5},
   \nonumber \\
   \tilde C^0_{\;\; 0ij}
   &=& L^{-2} {\cal O}^{-6},
   \nonumber \\
   \tilde C^0_{\;\; i0j}
   &=& {1 \over c^2} \left[
       {1 \over 2} \left( U + V \right)_{,i|j}
       - {1 \over 6} \Delta \left( U + V \right) \gamma_{ij} \right]
       + L^{-2} {\cal O}^{-4},
   \nonumber \\
   \tilde C^0_{\;\; ijk}
   &=& {1 \over c^3} {a \over 2} \left[
       \left( P^l_{\;\; |l[k}
       - \Delta P_{[k}
       - 2 K P_{[k}
       \right) \gamma_{j]i}
       - 2 P_{[j|k]i} \right]
       + L^{-2} {\cal O}^{-5},
   \nonumber \\
   \tilde C^i_{\;\; 00j}
   &=& {1 \over c^2} {1 \over a^2} \left[
       {1 \over 2} \left( U + V \right)^{|i}_{\;\;\; j}
       - {1 \over 6} \Delta \left( U + V \right) \delta^i_j
       \right]
       + L^{-2} {\cal O}^{-4},
   \nonumber \\
   \tilde C^i_{\;\; 0jk}
   &=& {1 \over c^3} {1 \over 2 a} \left[
       \left( P^l_{\;\; |l[k}
       - \Delta P_{[k}
       + 2 K P_{[k} \right) \delta^i_{j]}
       - 2 P_{[j|k]}^{\;\;\;\;\;\; i} \right]
       + L^{-2} {\cal O}^{-5},
   \nonumber \\
   \tilde C^i_{\;\; j0k}
   &=& {1 \over c^3} {1 \over 4 a} \left[
       \left( P^{l\;\;\; i}_{\;\; |l}
       - \Delta P^i
       + 2 K P^i \right) \gamma_{jk}
       - \left( P^l_{\;\; |lj}
       - \Delta P_j
       + 2 K P_j \right) \delta^i_k
       + 2 \left( P^i_{\;\; |j}
       - P_j^{\;\; |i} \right)_{|k} \right]
       + L^{-2} {\cal O}^{-5},
   \nonumber \\
   \tilde C^i_{\;\; jkl}
   &=& {1 \over c^2} \left[
       {2 \over 3} \Delta \left( U + V \right)
       \delta^i_{[k} \gamma_{l]j}
       + \left( U + V \right)_{,j|[k} \delta^i_{l]}
       - \left( U + V \right)^{,i}_{\;\;\; |[k} \gamma_{l]j}
       \right]
       + L^{-2} {\cal O}^{-4}.
\eea We can check the tracefree nature of the Weyl tensor: $\tilde
C^b_{\;\; bcd} = 0 = \tilde C^c_{\;\; bcd}$.  Here we encounter
non-vanishing traces involving $K$ terms, like \bea
   & & \tilde C^b_{\;\; b0i} = - {1 \over c^3} {K \over a} P_i
       + L^{-2} {\cal O}^{-5}, \quad
       \tilde C^c_{\;\; 0ci} = - {1 \over c^3} {2 K \over a} P_i
       + L^{-2} {\cal O}^{-5},
\eea whereas \bea
   & &
       \tilde C^c_{\;\; 0c0} = L^{-2} {\cal O}^{-4}, \quad
       \tilde C^b_{\;\; bij} = L^{-2} {\cal O}^{-4}, \quad
       \tilde C^c_{\;\; ic0} = L^{-2} {\cal O}^{-5}, \quad
       \tilde C^c_{\;\; icj} = L^{-2} {\cal O}^{-4}.
\eea This indicates that the presence of $K$ terms in the 1PN order
appears to be not reliable. Later this point will be resolved as we
show that the $K$ terms indeed can be related to the higher order
terms in the PN expansion, see below Eq.\ (\ref{K-issue}). Until we
reach such a conclusion about the effect of $K$ term we will keep it
in our equations.

\subsection{Energy-momentum tensor}

The normalized fluid four-vector $\tilde u^a$ with $\tilde u^a
\tilde u_a \equiv -1$ gives, to 1PN order, \bea
   \tilde u^0
   &=& 1
       + {1 \over c^2} \left( {1 \over 2} v^2 + U \right)
       + {1 \over c^4} \left[ {3 \over 8} v^4
       + v^2 \left( {3 \over 2} U + V \right)
       + {1 \over 2} U^2 + 2 \Phi - v^i P_i \right]
       + {\cal O}^{-6},
   \nonumber \\
   \tilde u^i
   &\equiv& {1 \over c} {1 \over a} v^i \tilde u^0,
   \nonumber \\
   \tilde u_0
   &=& - \left\{ 1
       + {1 \over c^2} \left( {1 \over 2} v^2 - U \right)
       + {1 \over c^4} \left[ {3 \over 8} v^4
       + v^2 \left( {1 \over 2} U + V \right)
       + {1 \over 2} U^2 - 2 \Phi \right] \right\}
       + {\cal O}^{-6},
   \nonumber \\
   \tilde u_i
   &=& a \Bigg\{ {1 \over c} v_i
       + {1 \over c^3} \left[
       v_i \left( {1 \over 2} v^2 + U + 2 V \right) - P_i
       \right]
       \Bigg\}
       + {\cal O}^{-5},
   \label{u-def}
\eea where the index of $v^i$ is based on $\gamma_{ij}$; $v^i$
corresponds to the peculiar velocity field. The energy-momentum
tensor is decomposed into fluid quantities as follows \bea
   & & \tilde T_{ab}
       = \tilde \varrho c^2 \left( 1 + {1 \over c^2} \tilde \Pi \right)
       \tilde u_a \tilde u_b
       + \tilde p \left( \tilde u_a \tilde u_b + \tilde g_{ab} \right)
       + \tilde q_a \tilde u_b + \tilde q_b \tilde u_a
       + \tilde \pi_{ab}.
   \label{T_ab}
\eea The energy density $\tilde \mu (\equiv \tilde \varrho c^2 +
\tilde \varrho \tilde \Pi)$ is decomposed into the material energy
density $\tilde \varrho c^2$ and the internal energy density $\tilde
\varrho \tilde \Pi$; $\tilde p$, $\tilde q_a$, and $\tilde \pi_{ab}$
are the isotropic pressure, the flux, and the anisotropic stress,
respectively. We have $\tilde q_a \tilde u^a \equiv 0$,
       $\tilde \pi_{ab} \tilde u^b \equiv 0$,
       $\tilde \pi^c_c \equiv 0$, and
       $\tilde \pi_{ab} \equiv \tilde \pi_{ba}$, thus \bea
   & & \tilde q_0 = - {1 \over c} {1 \over a} \tilde q_i v^i, \quad
       \tilde \pi_{0i} = - {1 \over c} {1 \over a} \tilde \pi_{ij} v^j, \quad
       \tilde \pi_{00} = {1 \over c^2} {1 \over a^2} \tilde \pi_{ij} v^i v^j.
\eea We introduce \bea
   & & \tilde q_i \equiv {1 \over c} a Q_i, \quad
       \tilde \pi_{ij} \equiv a^2 \Pi_{ij},
   \label{Q-def}
\eea where indices of $Q_i$ and $\Pi_{ij}$ are based on
$\gamma_{ij}$. We take $\tilde q_a$ and $\tilde \pi_{ab}$ to have
post-Newtonian orders as introduced in Eq.\ (\ref{Q-def}), see for
example \cite{Greenberg-1971}. This will be justified by the energy
conservation equation in the Newtonian context which will be derived
later, see Eq.\ (\ref{adiabatic-eq-c}). In a strictly single
component situation we can always remove $Q_i$ by following the
fluid element. However, there exist situations where we have the
additional flux terms present even when we follow the fluid
elements; the fundamental origin of such flux terms can be traced to
the presence of additional components in the energy-momentum. Thus,
in our case it is more general and convenient to keep the flux terms
separately. Up to the 1PN order the condition $\tilde \pi^c_c \equiv
0$ gives \bea
   & & \Pi^i_i = {1 \over c^2} \Pi_{ij} v^i v^j.
\eea We also set \bea
   & & \tilde \varrho \equiv \varrho, \quad
       \tilde \Pi \equiv \Pi, \quad
       \tilde p \equiv p.
\eea Dimensions are as follows \bea
   & & [\tilde u_a] = [\tilde u^a] = 1, \quad
       [\tilde T_{ab}] = [\tilde T^a_b] = [\tilde T^{ab}]
       = [\tilde p] = [\tilde q_a] = [\tilde q^a] = [\tilde \pi_{ab}]
       = [\tilde \varrho c^2] = M L^{-1} {\cal T}^{-2},
   \nonumber \\
   & &
       [v_i] = [v^i] = [c], \quad
       [\Pi] = [c^2], \quad
       [Q_i] = [Q^i] = [c \tilde q_a] = [\varrho c^3], \quad
       [\Pi_{ij}] = [\Pi^i_j] = [\Pi^{ij}] = [\tilde \pi_{ab}] = [\varrho c^2].
\eea We have \bea
   \tilde T_{00}
   &=& \varrho c^2 \Bigg\{
       1 + {1 \over c^2} \left( v^2 - 2 U + \Pi \right)
       + {1 \over c^4} \Bigg[ v^2 \left( v^2 + 2 V +
       \Pi + {p \over \varrho} \right)
       - 2 U \Pi + 2 U^2 - 4 \Phi
   \nonumber \\
   & &
       + {1 \over \varrho} \left( 2 Q_i v^i
       + \Pi_{ij} v^i v^j \right)
       \Bigg]
       + {\cal O}^{-6}
       \Bigg\},
   \nonumber \\
   \tilde T_{0i}
   &=& - a \varrho c^2 \left\{ {1 \over c} v_i
       + {1 \over c^3} \left[ v_i \left( v^2 + 2 V + \Pi
       + {p \over \varrho} \right)
       - P_i
       + {1 \over \varrho} \left( Q_i
       + \Pi_{ij} v^j \right) \right]
       + {\cal O}^{-5} \right\},
   \nonumber \\
   \tilde T_{ij}
   &=& a^2 \varrho c^2 \Bigg\{
       {1 \over c^2} \left( v_i v_j + {p \over \varrho}
       \gamma_{ij} + {1 \over \varrho} \Pi_{ij} \right)
   \nonumber \\
   & &
       + {1 \over c^4} \left[
       v_i v_j \left( v^2 + 2 U + 4 V + \Pi
       + {p \over \varrho} \right)
       + 2 V {p \over \varrho} \gamma_{ij}
       - 2 v_{(i} P_{j)}
       + {2 \over \varrho} Q_{(i} v_{j)}
       \right]
       + {\cal O}^{-6} \Bigg\},
   \nonumber \\
   \tilde T
   &=& - \varrho c^2 \left[ 1
       + {1 \over c^2} \left( \Pi - 3 {p \over \varrho} \right)
       + {\cal O}^{-6} \right].
   \label{Tab}
\eea {}From Eq.\ (\ref{T_ab}) covariantly we have $\tilde T = -
\tilde \varrho c^2 \left( 1 + {1 \over c^2} \tilde \Pi \right) + 3
\tilde p$.

\subsection{Fluid-frame kinematic quantities}

The projection tensor $\tilde h_{ab}$ based on the fluid-frame
four-vector $\tilde u^a$ is defined as \bea
   & & \tilde h_{ab} \equiv \tilde g_{ab} + \tilde u_a \tilde u_b.
\eea The kinematic quantities are defined as \cite{covariant} \bea
   & & \tilde \theta_{ab} \equiv \tilde h^c_{a} \tilde h^d_{b}
       \tilde u_{c;d}, \quad
       \tilde \theta \equiv \tilde u^a_{\;\; ;a}, \quad
       \tilde \sigma_{ab} \equiv \tilde \theta_{(ab)}
       - {1 \over 3} \tilde \theta \tilde h_{ab}, \quad
       \tilde \omega_{ab} \equiv \tilde \theta_{[ab]}, \quad
       \tilde a_a \equiv \tilde {\dot {\tilde u}}_a
       \equiv \tilde u_{a;b} \tilde u^b,
   \label{kinematic-quantities-def}
\eea where we have $\tilde u^a \tilde \theta_{ab} \equiv 0$, $\tilde
u^a \tilde \sigma_{ab} \equiv 0$, $\tilde u^a \tilde \omega_{ab}
\equiv 0$, $\tilde u^a \tilde a_a \equiv 0$, and $\tilde \theta
\equiv \tilde \theta^a_{\;\; a}$. The kinematic quantities $\tilde
\theta$, $\tilde a_a$, $\tilde \sigma_{ab}$, and $\tilde
\omega_{ab}$ are the expansion scalar, the acceleration vector, the
shear tensor, and the vorticity tensor, respectively.

To 1PN order, the projection tensor becomes \bea
   & & \tilde h^0_0
       = - {1 \over c^2} v^2
       - {1 \over c^4} \left[ v^2 \left( v^2 + 2 U + 2 V \right)
       - v^i P_i \right]
       + {\cal O}^{-6},
   \nonumber \\
   & & \tilde h^0_i
       = a \left\{ {1 \over c} v_i
       + {1 \over c^3} \left[ v_i \left( v^2 + 2 U + 2 V \right)
       - P_i \right] \right\}
       + {\cal O}^{-5},
   \nonumber \\
   & & \tilde h^i_0
       = - {1 \over c} {1 \over a} v^i \left\{
       1 + {1 \over c^2} v^2
       + {1 \over c^4} \left[ v^2 \left( v^2 + 2 U + 2 V \right)
       - v^j P_j \right] \right\}
       + {\cal O}^{-7},
   \nonumber \\
   & &\tilde h^i_j
       = \delta^i_j
       + {1 \over c^2} v^i v_j
       + {1 \over c^4} \left[ v^i v_j \left( v^2 + 2 U + 2 V
       \right) - v^i P_j \right]
       + {\cal O}^{-6}.
\eea The kinematic quantities become \bea
   \tilde \theta_{ij}
   &=& {1 \over c} a \left( v_{i|j}
       + \dot a \gamma_{ij} \right)
       + {1 \over c^3} a \Bigg\{
       v_j \left( a v_i \right)^\cdot
       + v_{i|j} \left( {1 \over 2} v^2 + U + 2 V \right)
       + v^k \left( v_i v_{k|j} + v_j v_{i|k} \right)
   \nonumber \\
   & &
       + a \left[ \dot V \
       + {\dot a \over a} \left( U + 2 V \right)
       + {1 \over a} v^k V_{,k}
       + {1 \over 2} {\dot a \over a} v^2 \right] \gamma_{ij}
       + 2 v_{[i} \left( U + V \right)_{,j]}
       - P_{[i|j]} \Bigg\}
       + L^{-1} {\cal O}^{-5},
   \\
   \tilde \theta
   &=& {1 \over c} \left( 3 {\dot a \over a}
       + {1 \over a} v^i_{\;\; |i} \right)
       + {1 \over c^3} \left[ 3 \dot V + 3 {\dot a \over a} U
       + {3 \over a} V_{,i} v^i
       + {1 \over a} U v^i_{\;\; |i}
       + {1 \over 2 a^3} \left( a^3 v^2 \right)^\cdot
       + {1 \over 2 a} \left( v^2 v^i \right)_{|i}
       \right]
       + L^{-1} {\cal O}^{-5},
   \\
   \tilde \sigma_{ij}
   &=& {1 \over c} a \left( v_{(i|j)}
       - {1 \over 3} v^k_{\;\; |k} \gamma_{ij} \right)
       + {1 \over c^3} a \Bigg\{
       v_{(i} \left( a v_{j)} \right)^\cdot
       + v_{(i|j)} \left( {1 \over 2} v^2 + U + 2 V \right)
       + v^k \left( v_{k|(i} v_{j)}
       + v_{(i} v_{j)|k} \right)
   \nonumber \\
   & &
       - v_i v_j \left( \dot a
       + {1 \over 3} v^k_{\;\; |k} \right)
       - {1 \over 3} \gamma_{ij}
       \left[ {1 \over 2} a \left( v^2 \right)^\cdot
       + \left( U + 2 V \right) v^k_{\;\; |k}
       + {1 \over 2} \left( v^2 v^k \right)_{|k} \right]
       \Bigg\}
       + L^{-1} {\cal O}^{-5},
   \\
   \tilde \omega_{ij}
   &=& {1 \over c} a v_{[i|j]}
       + {1 \over c^3} a \left\{
       v_{[j} \left( a v_{i]} \right)^\cdot
       + v_{[i|j]} \left( {1 \over 2} v^2 + U + 2 V \right)
       + v^k \left( v_{k|[j} v_{i]}
       + v_{[j} v_{i]|k} \right)
       + 2 v_{[i} \left( U + V \right)_{,j]}
       - P_{[i|j]} \right\}
   \nonumber \\
   & &
       + L^{-1} {\cal O}^{-5},
   \\
   \tilde a_i
   &=& - {1 \over c^2} U_{,i} \left( 1 + {1 \over c^2} v^2
       \right)
       - {1 \over c^4} 2 \Phi_{,i}
       + \left[ 1 + {1 \over c^2} \left( {1 \over 2} v^2
       + U \right) \right] \left( {\partial \over \partial t}
       + {1 \over a} {\bf v} \cdot \nabla \right) {1 \over c} \tilde u_i
       + {1 \over c^4} \left( - v^2 V_{,i}
       + v_j P^j_{\;\; |i} \right)
   \nonumber \\
   & & \qquad
       + L^{-1} {\cal O}^{-6}.
\eea From $\tilde u^c \tilde a_c \equiv 0$ and $\tilde u^b \tilde
\theta_{ab} \equiv 0$ we have \bea
   & & \tilde a_0 = - {1 \over c} {1 \over a} v^i \tilde a_i, \quad
       \tilde \theta_{0i} = - {1 \over c} {1 \over a} v^j \tilde \theta_{ij},
       \quad
       \tilde \theta_{00}
       = {1 \over c^2} {1 \over a^2} v^i v^j \tilde \theta_{ij},
\eea and similarly for $\tilde \sigma_{0a}$ and $\tilde
\omega_{0a}$.

The electric and the magnetic parts of the Weyl (conformal) tensor
are introduced as \bea
   & & \tilde E_{ab} \equiv \tilde C_{acbd} \tilde u^c \tilde u^d,
       \quad
       \tilde H_{ab} \equiv {1 \over 2}
       \tilde \eta_{ac}^{\;\;\;\; ef} \tilde C_{efbd} \tilde u^c
       \tilde u^d,
   \label{Weyl-tensor-def}
\eea where $\tilde \eta^{abcd} \equiv {1 \over \sqrt{-\tilde g}}
\epsilon^{abcd}$ is the totally antisymmetric tensor density with
$\epsilon^{abcd}$, the totally antisymmetric Levi-Civita symbol with
$\epsilon^{0123} \equiv 1$. $\tilde E_{ab}$ and $\tilde H_{ab}$ are
symmetric, tracefree and $\tilde u^a \tilde E_{ab} = 0 = \tilde u^a
\tilde H_{ab}$. To 1PN order we have \bea
   \tilde E^i_j
   &=& - \tilde C^i_{\;\; 00j}
   \nonumber \\
   &=& - {1 \over c^2} {1 \over a^2}
       \left[ {1 \over 2} \left( U + V \right)^{|i}_{\;\;\; j}
       - {1 \over 6} \Delta \left( U + V \right) \delta^i_j \right]
       + L^{-2} {\cal O}^{-4},
   \\
   \tilde H^i_j
   &=& - {1 \over 2} \tilde u^0 \tilde u^0
       \left[ \tilde \eta^{i \;\;\;\; k}_{\;\; 0l}
       \left( \tilde C^l_{\;\; k0j}
       - {1 \over c} {1 \over a} v^l \tilde C^0_{\;\; k0j} \right)
       + \tilde \eta^{i \;\;\;\; 0}_{\;\;lk}
       {1 \over c} {1 \over a} v^l \tilde C^k_{\;\; 00j} \right]
   \nonumber \\
   &=& {1 \over c^3} {1 \over 2 a^3} \eta^{ikl}
       \left\{
       \left[ {1 \over 2} \left( P^{m}_{\;\;\;|ml}
       - \Delta P_l + 2 K P_l \right)
       + {1 \over 3} v_l \Delta \left( U + V \right) \right]
       \gamma_{kj}
       + P_{l|kj}
       - v_l \left( U + V \right)_{|kj} \right\}
       + L^{-2} {\cal O}^{-5},
   \label{H-fluid-frame}
\eea where we introduced $\eta^{ijk} \equiv {1 \over \sqrt{\gamma}}
\epsilon^{ijk}$ with $\epsilon^{ijk}$, the totally antisymmetric
Levi-Civita symbol with $\epsilon^{123} \equiv 1$. Thus $\tilde
\eta^{0ijk} = \sqrt{ \gamma \over - \tilde g } \eta^{ijk}$. Indices
of $\eta^{ijk}$ are based on $\gamma_{ij}$. $\tilde E_{0a}$ follows
from $\tilde u^b \tilde E_{ab} = 0$ which gives $\tilde E_{0a} = -
(\tilde u^i / \tilde u^0) \tilde E_{ia} = - (v^i/c a) \tilde
E_{ia}$, thus $\tilde E_{0i} \sim L^{-2} {\cal O}^{-3}$ and $\tilde
E_{00} \sim L^{-2} {\cal O}^{-4}$. {}For nonvanishing $K$, $\tilde
H_{ab}$ is not tracefree to $L^{-2} {\cal O}^{-3}$, but as we
mentioned earlier the term involving $K$ is already ${\cal O}^{-2}$
order higher, thus consistent.

\subsection{Normal-frame kinematic quantities}

The normalized normal-frame four-vector $\tilde n^a$, which is
normal to the space-like hypersurfaces, is defined as $\tilde n_i
\equiv 0$ with $\tilde n^a \tilde n_a \equiv -1$. To 1PN order we
have \bea
   & & \tilde n^0 = 1 + {1 \over c^2} U
       + {1 \over c^4} \left( {1 \over 2} U^2 + 2 \Phi \right) + {\cal
       O}^{-6}, \quad
       \tilde n^i = {1 \over c^3} {1 \over a} P^i + {\cal O}^{-5},
   \nonumber \\
   & & \tilde n_0 = -1 + {1 \over c^2} U
       - {1 \over c^4} \left( {1 \over 2} U^2
       - 2 \Phi \right) + {\cal O}^{-6}, \quad
       \tilde n_i \equiv 0.
\eea By setting $\tilde u_i \equiv 0$ in Eq.\ (\ref{u-def}) we also
recover the normal vector, thus \bea
   & & v_i = {1 \over c^2} P_i + L {\cal T}^{-1} {\cal O}^{-4},
\eea where $v_i$ is, say, the velocity of the normal frame vector.
The projection tensor based on $\tilde n^a$ becomes \bea
   & & \tilde h_{ij} = \tilde g_{ij}, \quad
       \tilde h_{0i} = \tilde g_{0i}, \quad
       \tilde h_{00} = {\cal O}^{-6}, \quad
       \tilde h^i_j = \delta^i_j, \quad
       \tilde h^i_0 = - {1 \over c^3} {1 \over a} P^i + {\cal O}^{-5}, \quad
       \tilde h^0_i = 0 = \tilde h^0_0.
\eea Kinematic quantities based on $\tilde n^a$ become \bea
   \tilde \theta
   &=& {1 \over c} 3 {\dot a \over a}
       + {1 \over c^3} \left( 3 \dot V + 3 {\dot a \over a} U
       + {1 \over a} P^i_{\;\; |i} \right)
       + L^{-1} {\cal O}^{-5},
   \label{theta-n-frame} \\
   \tilde \sigma_{ij}
   &=& {1 \over c^3} a \left( P_{(i|j)}
       - {1 \over 3} P^k_{\;\; |k} \gamma_{ij} \right)
       + L^{-1} {\cal O}^{-5},
   \label{shear-n-frame} \\
   \tilde \omega_{ij}
   &=& 0,
   \\
   \tilde a_i
   &=& - {1 \over c^2} U_{,i}
       - {1 \over c^4} 2 \Phi_{,i}
       + L^{-1} {\cal O}^{-6}.
\eea $\tilde \sigma_{0c}$ follows from $\tilde \sigma_{ac} \tilde
n^c \equiv 0$, thus $\tilde \sigma_{0i} \sim L^{-1} {\cal O}^{-6}$
and $\tilde \sigma_{00} \sim L^{-1} {\cal O}^{-9}$; we have $\tilde
\omega_{ab} = 0$. Similarly, $\tilde a_0$ follows from $\tilde a_c
\tilde n^c \equiv 0$, thus $\tilde a_0 \sim L^{-1} {\cal O}^{-5}$.

The electric and the magnetic parts of Weyl tensor based on $\tilde
n^a$ give \bea
   \tilde E^i_j
   &=& - \tilde C^i_{\;\; 00j}
   \nonumber \\
   &=& - {1 \over c^2} {1 \over a^2}
       \left[ {1 \over 2} \left( U + V \right)^{|i}_{\;\;\; j}
       - {1 \over 6} \Delta \left( U + V \right) \delta^i_j \right]
       + L^{-2} {\cal O}^{-4},
   \\
   \tilde H^i_j
   &=& {1 \over 2} \tilde n_0 \tilde n_0
       \tilde \eta^{0ikl} \tilde C^0_{\;\; jkl}
   \nonumber \\
   &=& {1 \over c^3} {1 \over 2 a^3} \eta^{ikl}
       \left[ {1 \over 2} \left( P^{m}_{\;\;\;|ml}
       - \Delta P_l + 2 K P_l \right) \gamma_{jk}
       + P_{l|kj} \right]
       + L^{-2} {\cal O}^{-5}.
\eea Thus, $\tilde E^i_j$ is the same in both frames to the 1PN
order; compared with Eq.\ (\ref{H-fluid-frame}) $\tilde H^i_j$
differs. The nonvanishing $K$ term which causes $\tilde H_{ab}$ to
be not tracefree to $L^{-2} {\cal O}^{-3}$ can be ignored because
the $K$ term is of the ${\cal O}^{-2}$ order.

\subsection{ADM quantities}
                                       \label{sec:ADM-quantities}

In the Arnowitt-Deser-Misner (ADM) notation the metric and fluid
quantities are \cite{ADM} \bea
   & & \tilde g_{00} \equiv - N^2 + N^i N_i, \quad
       \tilde g_{0i} \equiv N_i, \quad
       \tilde g_{ij} \equiv h_{ij},
   \label{ADM-metric-def} \\
   & & \tilde n_0 \equiv - N, \quad
       \tilde n_i \equiv 0, \quad
       \tilde n^0 = N^{-1}, \quad
       \tilde n^i = - N^{-1} N^i,
   \label{n_a-def} \\
   & & E \equiv \tilde n_a \tilde n_b \tilde T^{ab}, \quad
       J_i \equiv - \tilde n_b \tilde T^b_i, \quad
       S_{ij} \equiv \tilde T_{ij}, \quad
       S \equiv h^{ij} S_{ij}, \quad
       \bar S_{ij} \equiv S_{ij}
       - {1\over 3} h_{ij} S,
   \label{ADM-fluid-def}
\eea where $N_i$, $J_i$ and $S_{ij}$ are based on $h_{ij}$ as the
metric, and $h^{ij}$ is the inverse metric of $h_{ij}$. The
extrinsic curvature is \bea
   & & K_{ij} \equiv {1\over 2N} \left( N_{i:j}
       + N_{j:i} - h_{ij,0} \right), \quad
       \bar K \equiv h^{ij} K_{ij}, \quad
       \bar K_{ij} \equiv K_{ij}
       - {1\over 3} h_{ij} \bar K,
   \label{extrinsic-curvature-def}
\eea where indices of $K_{ij}$ are based on $h_{ij}$; a colon `$:$'
denotes the covariant derivative based on $h_{ij}$ with
$\Gamma^{(h)i}_{\;\;\;\;\;jk} \equiv {1 \over 2} h^{il} \left(
h_{jl,k} + h_{lk,j} - h_{jk,l} \right)$. The intrinsic curvatures
are based on the metric $h_{ij}$ \bea
   & & R^{(h)i}_{\;\;\;\;\;\;jkl}
       \equiv
       \Gamma^{(h)i}_{\;\;\;\;\;jl,k}
       - \Gamma^{(h)i}_{\;\;\;\;\;jk,l}
       + \Gamma^{(h)m}_{\;\;\;\;\;jl}
       \Gamma^{(h)i}_{\;\;\;\;\;km}
       - \Gamma^{(h)m}_{\;\;\;\;\;jk}
       \Gamma^{(h)i}_{\;\;\;\;\;lm},
   \nonumber \\
   & & R^{(h)}_{ij}
       \equiv R^{(h)k}_{\;\;\;\;\;\;\;ikj}, \quad
       R^{(h)} \equiv h^{ij} R^{(h)}_{ij}, \quad
       \bar R^{(h)}_{ij} \equiv R^{(h)}_{ij}
       - {1\over 3} h_{ij} R^{(h)}.
   \label{ADM-curvature}
\eea A complete set of ADM equations will be presented in Sec.\
\ref{sec:ADM-eqs}.

To the 1PN order we have \bea
   & & N = 1 - { 1\over c^2} U
       + { 1\over c^4} \left( {1 \over 2} U^2 - 2 \Phi \right)
       + {\cal O}^{-6},
   \nonumber \\
   & & N^i = - {1 \over c^3} {1 \over a} P^i
       \left( 1 - {1 \over c^2} U \right)
       + {\cal O}^{-7}, \quad
       N_i = - {1 \over c^3} a P_i \left[
       1 + {1 \over c^2} \left( 2 V - U \right) \right]
       + {\cal O}^{-7},
   \nonumber \\
   & & h_{ij} = a^2 \left( 1 + {1 \over c^2} 2 V \right)
       \gamma_{ij}
       + {\cal O}^{-4}, \quad
       h^{ij} = {1 \over a^2} \left( 1 - {1 \over c^2} 2 V \right)
       \gamma^{ij}
       + {\cal O}^{-4},
   \\
   & & \Gamma^{(h)i}_{\;\;\;\;\;jk}
       = \Gamma^{(\gamma)i}_{\;\;\;\;\;jk}
       + {1 \over c^2} \left( V_{,k} \delta^i_j
       + V_{,j} \delta^i_k
       - V^{,i} \gamma_{jk} \right)
       + L^{-1} {\cal O}^{-4},
   \\
   & & R^{(h)}_{ij}
       = R^{(\gamma)}_{ij}
       - {1 \over c^2} \left( V_{,i|j}
       + \gamma_{ij} \Delta V \right)
       + L^{-2} {\cal O}^{-4}, \quad
       R^{(h)}
       = {1 \over a^2} \left[ 6 K
       - {1 \over c^2} 4 \left( \Delta + 3 K \right) V \right]
       + L^{-2} {\cal O}^{-4},
   \\
   & & K_{ij}
       = - {1 \over c} a \dot a \gamma_{ij}
       - {1 \over c^3} a^2 \left\{
       \left[ \dot V + {\dot a \over a} \left( U + 2 V \right)
       \right] \gamma_{ij}
       + {1 \over a} P_{(i|j)} \right\}
       + L^{-1} {\cal O}^{-5},
   \nonumber \\
   & & \bar K
       = - {1 \over c} 3 {\dot a \over a}
       - {1 \over c^3} \left[ 3 \left( \dot V + {\dot a \over a} U
       \right)
       + {1 \over a} P^i_{\;\; |i} \right]
       + L^{-1} {\cal O}^{-5},
   \label{ADM-extrinsic-curvature} \\
   & & E = \varrho c^2 \Bigg\{
       1 + { 1\over c^2} \left( v^2 + \Pi \right)
       + {1 \over c^4} \left[ v^2 \left( v^2
       + 2 U + 2 V + \Pi + {p \over \varrho} \right)
       + 2 U \Pi - 2 P_i v^i
       + {1 \over \varrho} \left( 2 Q_i v^i
       + \Pi_{ij} v^i v^j \right) \right]
       + {\cal O}^{-6} \Bigg\},
   \nonumber \\
   & & J_i = a \varrho c^2 \left\{
       {1 \over c} v_i
       + {1 \over c^3} \left[ v_i \left( v^2 + U + 2 V + \Pi
       + {p \over \varrho} \right) - P_i
       + {1 \over \varrho} \left( Q_i
       + \Pi_{ij} v^j \right) \right]
       + {\cal O}^{-5} \right\},
   \nonumber \\
   & & S_{ij} = a^2 \varrho c^2 \Bigg\{
       {1 \over c^2} \left( v_i v_j
       + {p \over \varrho} \gamma_{ij}
       + {1 \over \varrho} \Pi_{ij} \right)
   \nonumber \\
   & & \qquad
       + {1 \over c^4} \left[
       v_i v_j \left( v^2 + 2 U + 4 V + \Pi
       + {p \over \varrho} \right)
       + 2 {p \over \varrho} V \gamma_{ij}
       - 2 v_{(i} P_{j)}
       + {2 \over \varrho} Q_{(i} v_{j)} \right]
       + {\cal O}^{-6} \Bigg\},
   \nonumber \\
   & & S = \varrho c^2 \left\{
       {1 \over c^2} \left( v^2 + 3 {p \over \varrho} \right)
       + {1 \over c^4} \left[ v^2 \left( v^2 + 2 U + 2 V + \Pi
       + {p \over \varrho} \right)
       - 2 P_i v^i + {1 \over \varrho} \left( 2 Q_i v^i
       + \Pi_{ij} v^i v^j \right) \right]
       + {\cal O}^{-6} \right\}.
   \label{ADM-fluid}
\eea

%
%
\section{Derivations}
                                         \label{sec:Derivations}

\subsection{Equations of motion}
                                                 \label{sec:EOM}

Using Eq.\ (\ref{Tab}) the energy and momentum conservation
equations give \bea
   0
   &=& - { 1\over c} \tilde T^b_{0;b}
   \nonumber \\
   &=& {1 \over a^3} \left( a^3 \sigma \right)^\cdot
       + {1 \over a} \left[ \sigma v^i
       + {1 \over c^2} \left( Q^i
       + \Pi^i_j v^j \right)
       \right]_{|i}
       + {1 \over c^2} \varrho \left[ \dot V
       + {1 \over a} v^i \left( V - U \right)_{,i}
       + {\dot a \over a} v^2
       - {\dot p \over \varrho} \right]
       + \varrho {\cal T}^{-1} {\cal O}^{-4},
   \label{E-conserv-0} \\
   0
   &=& {1 \over a} \tilde T^b_{i;b}
   \nonumber \\
   &=& {1 \over a^4} \left\{ a^4 \left[ \sigma v_i
       + {1 \over c^2} \left( Q_i + \Pi_{ij} v^j \right)
       \right] \right\}^\cdot
       + {1 \over a} \left\{ \sigma v_i v^j
       + \Pi^j_i
       + {1 \over c^2} \left[
       Q^j v_i + Q_i v^j
       - 2 \left( U + V \right) \Pi^j_i \right]
       \right\}_{|j}
   \nonumber \\
   & &
       + {1 \over a} \left( 1 - {1 \over c^2} 2 U \right) p_{,i}
       - {1 \over a} \left( \sigma U_{,i}
       + {1 \over c^2} \varrho v^2 V_{,i} \right)
   \nonumber \\
   & &
       + {1 \over c^2} \varrho \left\{
       v_i \left( {\partial \over \partial t}
       + {1 \over a} {\bf v} \cdot \nabla \right) \left( U + 3 V
       \right)
       + {2 \over a} \left[ \left( U + V \right) U_{,i} - \Phi_{,i}
       \right]
       - {1 \over a} \left( a P_i \right)^\cdot
       - {2 \over a} v^j P_{[i|j]}
       + {1 \over \varrho a} \left( U + 3 V \right)_{,j} \Pi^j_i
       \right\}
   \nonumber \\
   & &
       + \varrho L {\cal T}^{-2} {\cal O}^{-4},
   \label{Mom-conserv-0}
\eea where \bea
   & & \sigma \equiv \varrho \left[ 1
       + {1 \over c^2} \left( v^2 + 2 V + \Pi + {p \over
       \varrho} \right)
       \right].
\eea For $Q_i = 0 = \Pi_{ij}$, $V = U$, $a = 1$ and $\gamma_{ij} =
\delta_{ij}$ these equations reduce to Eqs.\ (64), (67) in
Chandrasekhar \cite{Chandrasekhar-1965}.

Equation (\ref{E-conserv-0}) can be written in another form as \bea
   & & {1 \over a^3} \left( a^3 \varrho^* \right)^\cdot
       + {1 \over a} \left( \varrho^* v^i \right)_{|i}
       + {1 \over c^2} \left[
       {1 \over a} \left( Q^i_{\;\;|i}
       + \Pi^i_j v^j_{\;\;|i} \right)
       + \varrho \left( {\partial \over \partial t}
       + {1 \over a} {\bf v} \cdot \nabla \right) \Pi
       + \left( 3 {\dot a \over a}
       + {1 \over a} \nabla \cdot {\bf v} \right) p
       \right]
       + \varrho {\cal T}^{-1} {\cal O}^{-4} = 0,
   \nonumber \\
   \label{E-conserv-00}
\eea where \bea
   & & \varrho^*
       \equiv \varrho {\sqrt{- \tilde g} \over a^3 \sqrt{\gamma}} \tilde u^0
       \equiv \varrho \left[ 1 + {1 \over c^2}
       \left( {1 \over 2} v^2 + 3 V \right) + {\cal O}^{-4} \right],
\eea with $a^3 \sqrt{\gamma}$ the background part of $\sqrt{- \tilde
g}$. This corresponds to Eq.\ (117) in Chandrasekhar
\cite{Chandrasekhar-1965}; see also Eq.\ (44) in
\cite{Chandrasekhar-1969}. The mass conservation (continuity)
equation $0 = \left( \tilde \varrho \tilde u^c \right)_{;c} = \tilde
{\dot {\tilde \varrho}} + \tilde \theta \tilde \varrho$ gives \bea
   0
   &=& c \left( \tilde \varrho \tilde u^c \right)_{;c}
       = c {1 \over \sqrt{-\tilde g}}
       \left( \varrho \sqrt{-\tilde g} \tilde u^c \right)_{,c}
   \nonumber \\
   &=& {1 \over a^3} \left( a^3 \varrho^* \right)^\cdot
       + {1 \over a} \left( \varrho^* v^i \right)_{|i}
       + \varrho {\cal T}^{-1} {\cal O}^{-4}
   \nonumber \\
   &=& {1 \over a^3} \left( a^3 \varrho \right)^\cdot
       + {1 \over a} \left( \varrho v^i \right)_{|i}
       + {1 \over c^2} \varrho \left( {\partial \over \partial t}
       + {1 \over a} {\bf v} \cdot \nabla \right)
       \left( {1 \over 2} v^2 + 3 V \right)
       + \varrho {\cal T}^{-1} {\cal O}^{-4}.
   \label{mass-conservation}
\eea Thus, if we {\it assume} the mass conservation, Eq.\
(\ref{E-conserv-00}) gives \bea
   & & \left( {\partial \over \partial t}
       + {1 \over a} {\bf v} \cdot \nabla \right) \Pi
       + \left( 3 {\dot a \over a} + {1 \over a} \nabla \cdot {\bf v} \right)
       {p \over \varrho}
       = - {1 \over \varrho a} \left( Q^i_{\;\;|i}
       + \Pi^i_j v^j_{\;\;|i}
       \right)
       + {\cal T}^{-1} {\cal O}^{-2}.
   \label{adiabatic-case}
\eea The specific entropy is introduced as $\tilde T d \tilde S = d
\left( \tilde \Pi / c^2 \right) + \tilde p / c^2 d \left(1/\tilde
\varrho \right)$, thus along the flow we have \bea
   & & \tilde T \tilde {\dot {\tilde S}}
       = {1 \over c^2} \left[ \tilde {\dot {\tilde \Pi}}
       + \tilde p \left( {1 / \tilde \varrho}
       \right)^{\tilde \cdot} \right]
       = {1 \over c^3} \left[ \left( {\partial \over \partial t}
       + {1 \over a} {\bf v} \cdot \nabla \right) \tilde \Pi
       + \left( 3 {\dot a \over a} + {1 \over a} \nabla \cdot {\bf v} \right)
       {\tilde p \over \tilde \varrho}
       + {\cal T}^{-1} {\cal O}^{-2}
       \right].
   \label{entropy-eq}
\eea This shows that for an adiabatic fluid flow the LHS of Eq.\
(\ref{adiabatic-case}) vanishes; this also makes the RHS to vanish,
which is naturally required for an ideal fluid. According to
Chandrasekhar \cite{Chandrasekhar-1969}: ``{\sl the conservation of
mass and the conservation of entropy are not independent
requirements in the framework of general relativity.} And the reason
for their {\sl independence} in the Newtonian limit is that in this
limit (``$c^2 \rightarrow \infty$'') [our Eq.\
(\ref{mass-conservation})] reduces simply to the equation of
continuity.''

In \cite{Blanchet-etal-1990} Blanchet, et al.\ suggested to use the
following new combination of variable instead of $v_i$ \bea
   & & v^*_i \equiv {1 \over c} {\sqrt{-\tilde g} \over a^4 \sqrt{\gamma}}
       {1 \over \varrho^*} \tilde T^0_i,
\eea which becomes \bea
   & & v^*_i = v_i
       + {1 \over c^2} \left[ v_i \left( {1 \over 2} v^2 + U + 2 V +
       \Pi + {p \over \varrho} \right)
       - P_i
       + {1 \over \varrho} \left( Q_i
       + \Pi_{ij} v^j \right) \right]
       + L {\cal T}^{-1} {\cal O}^{-4}.
\eea Notice that $v^*_i$ is directly related to the ADM flux vector
in Eq.\ (\ref{ADM-fluid}) \bea
   & & J_i = a \varrho c v^*_i.
   \label{J_i-v^*}
\eea Equation $\tilde T^b_{i;b} = 0$ can be written as follows \bea
   & & {1 \over c} \left( {\sqrt{-\tilde g} \over \sqrt{\gamma}}
       \tilde T^0_i \right)^\cdot
       + \left( {\sqrt{-\tilde g} \over \sqrt{\gamma}}
       \tilde T^j_i \right)_{|j}
       = {1 \over 2} {\sqrt{-\tilde g} \over \sqrt{\gamma}}
       \tilde T^{ab} \tilde g_{ab|i}
   \nonumber \\
   & & \qquad
       = {\sqrt{-\tilde g} \over \sqrt{\gamma}} \varrho \left\{
       U_{,i}
       + { 1\over c^2} \left[ U_{,i} \left( v^2 + \Pi \right)
       + V_{,i} \left( v^2 + 3 {p \over \varrho} \right)
       - v^j P_{j|i}
       + 2 \Phi_{,i} \right] + L^2 {\cal T}^{-2}  {\cal O}^{-4}\right\},
   \label{Blanchet-eq}
\eea and we have \bea
   & & {\sqrt{-\tilde g} \over \sqrt{\gamma}} \tilde T^j_i
       = a^3 v^j \varrho^* v^*_i
       + {\sqrt{-\tilde g} \over \sqrt{\gamma}}
       \left[ p \delta^j_i
       + \Pi^j_i
       + {1 \over c^2} \left( Q^j v_i
       - 2 V \Pi^j_i
       - \Pi_{ik} v^j v^k \right)
       + L {\cal T}^{-1} {\cal O}^{-4} \right].
\eea Thus, Eq.\ (\ref{Blanchet-eq}) can be arranged in the following
form \bea
   & & {1 \over a} \left( a v^*_i \right)^\cdot
       + {1 \over a} v^*_{i|j} v^j
       = {1 \over a} \left\{
       U_{,i}
       + {1 \over c^2} \left[ U_{,i} \left( {1 \over 2} v^2
       - U + \Pi \right)
       + V_{,i} \left( v^2 + 3 {p \over \varrho} \right)
       - v^j P_{j|i}
       + 2 \Phi_{,i} \right] \right\}
   \nonumber \\
   & & \qquad
       - {1 \over \varrho^* a}
       \left\{ \left[ 1 + {1 \over c^2} \left( 3 V - U \right) \right]
       \left[ p \delta^j_i
       + \Pi^j_i
       + {1 \over c^2} \left( Q^j v_i
       - 2 V \Pi^j_i
       - \Pi_{ik} v^j v^k \right) \right] \right\}_{|j}
   \nonumber \\
   & & \qquad
       + {1 \over c^2} {v^*_i \over \varrho^*}
       \left[ {1 \over a} \left( Q^j + \Pi^j_k v^k \right)_{|j}
       + \varrho \left( {\partial \over \partial t}
       + {1 \over a} {\bf v} \cdot \nabla \right) \Pi
       + \left( 3 {\dot a \over a}
       + {1 \over a} \nabla \cdot {\bf v} \right) p \right]
       + L {\cal T}^{-2} {\cal O}^{-4},
   \label{v^*_i-eq}
\eea where the last line vanishes in an adiabatic ideal fluid.

To the ${\cal O}^{-0}$ order, Eqs.\ (\ref{E-conserv-0}),
(\ref{Mom-conserv-0}) give the Newtonian mass and momentum
conservation equations, respectively \bea
   & &
       {1 \over a^3} \left( a^3 \varrho \right)^\cdot
       + {1 \over a} \nabla_i \left( \varrho v^i \right) = 0,
   \label{Newtonian-mass-conservation} \\
   & &
       {1 \over a} \left( a v_i \right)^\cdot
       + {1 \over a} v^j \nabla_j v_i
       + {1 \over a \varrho} \left( \nabla_i p
       + \nabla_j \Pi^j_i \right)
       - {1 \over a} \nabla_i U = 0.
   \label{Newtonian-momentum-conservation}
\eea Thus, we naturally have the Newtonian equations to 0PN order,
e.g., see \cite{Kofman-etal}. This can be compared with the
situation in perturbation approach where the Newtonian
correspondence can be achieved only after suitable choice of
different gauges for different variables, thus being a non-trivial
result even to the linear order in perturbation approach
\cite{Bardeen-1980,HN-Newtonian-1999}. Equations
(\ref{Newtonian-mass-conservation}),
(\ref{Newtonian-momentum-conservation}), as well as all the 1PN
equations in this section, are fully nonlinear.

In the Friedmann background we set the PN variables $U$, $\Phi$,
$P_i$, $V$, $v_i$, $Q_i$, and $\Pi_{ij}$ equal to zero, and set
$\varrho = \varrho_b$, $\Pi = \Pi_b$, and $p = p_b$, with $\mu
\equiv \varrho \left( c^2 + \Pi \right)$ and $\sigma_b \equiv
\varrho_b \left[ 1 + {1 \over c^2} \left( \Pi_b + p_b/\varrho_b
\right) \right]$. Equations (\ref{E-conserv-0}),
(\ref{mass-conservation}) give \bea
   & &
       \dot \mu_b + 3 {\dot a \over a}
       \left( \mu_b + p_b \right) = 0,
   \label{BG-eq1} \\
   & &
       \dot \varrho_b + 3 {\dot a \over a} \varrho_b = 0.
   \label{BG-eq1-1}
\eea By subtracting the background part, Eq.\ (\ref{E-conserv-0})
becomes \bea
   & &
       {1 \over a^3} \left[ a^3 \left( \sigma - \sigma_b \right) \right]^\cdot
       + {1 \over a} \left[ \sigma v^i
       + {1 \over c^2} \left( Q^i
       + \Pi^i_j v^j \right)
       \right]_{|i}
       + {1 \over c^2} \varrho \left[ \dot V
       + {1 \over a} v^i \left( V - U \right)_{,i}
       + {\dot a \over a} v^2
       - {\dot p - \dot p_b \over \varrho} \right]
       + \varrho {\cal T}^{-1} {\cal O}^{-4}
       = 0.
   \nonumber \\
   \label{E-conserv-pert-0}
\eea

\subsection{Einstein's equations}
                                           \label{sec:Einstein-eq}

We take Einstein's equations in the form \bea
   & & \tilde R^a_b = {8 \pi G \over c^4}
       \left( \tilde T^a_b - {1 \over 2} \tilde T \delta^a_b \right)
       + \Lambda \delta^a_b.
   \label{Einstein-eq}
\eea Dimensions are as follows \bea
   & & [\tilde g_{ab}] = 1, \quad
       [\tilde R_{ab}] = [\tilde R^a_b] = [\tilde R^{ab}] = L^{-2}, \quad
       [\tilde T_{ab}] = [\tilde T] = [\varrho c^2] = M L^{-1} {\cal T}^{-2}, \quad
       [\Lambda] = L^{-2}, \quad
       [G\varrho] = {\cal T}^{-2},
\eea where $M$ indicates the dimension of mass. To 1PN order, Eqs.\
(\ref{Rab}), (\ref{Tab}) give \bea
   & & - \Lambda
       + {1 \over c^2} \left( 3 {\ddot a \over a}
       + {\Delta \over a^2} U + 4 \pi G \varrho \right)
       + {1 \over c^4} \Bigg\{
       3 \ddot V
       + 3 {\dot a \over a} \left( \dot U + 2 \dot V \right)
       + 6 {\ddot a \over a} U
   \nonumber \\
   & & \qquad
       - {1 \over a^2} \left[ U^{,i} \left( U - V \right)_{,i}
       + 2 V \Delta U
       - 2 \Delta \Phi
       - \left( a P^i_{\;\;|i} \right)^\cdot \right]
       + 8 \pi G \varrho \left(
       v^2
       + {1 \over 2} \Pi
       + {3 \over 2} {p \over \varrho} \right)
       \Bigg\}
       = 0,
   \label{R_00-eq-raw} \\
   & & {1 \over c^3} \left[ 2 \left( \dot V + {\dot a \over a} U
       \right)_{,i}
       + {1 \over 2 a} \left( P^j_{\;\;|ji} - \Delta P_i
       - 2 K P_i \right)
       - 8 \pi G \varrho a v_i
       \right]
       = 0,
   \label{R_0i-eq-raw} \\
   & & \left( {2 K \over a^2} - \Lambda \right) \delta^i_j
       + {1 \over c^2} \left[
       \left( {\ddot a \over a} + 2 {\dot a^2 \over a^2}
       - {\Delta + 4 K \over a^2} V
       - 4 \pi G \varrho \right) \delta^i_j
       + {1 \over a^2} \left( U - V \right)^{|i}_{\;\;\; j} \right]
   \nonumber \\
   & & \qquad
       =
       {1 \over c^4} 8 \pi G \varrho \left[ v^i v_j
       + {1 \over 2} \left( \Pi - {p \over \varrho} \right) \delta^i_j
       + {1 \over \varrho} \Pi^i_j \right],
   \label{R_ij-eq-raw}
\eea where Eqs.\ (\ref{R_00-eq-raw}), (\ref{R_0i-eq-raw}), and
(\ref{R_ij-eq-raw}) are $\tilde R^0_0$, $\tilde R^0_i$, and $\tilde
R^i_j$ parts of Eq.\ (\ref{Einstein-eq}), respectively.

We kept the ${\cal O}^{-4}$ term on the RHS of Eq.\
(\ref{R_ij-eq-raw}) in order to get the correct Friedmann background
equation. To be consistent, we also need to keep the LHS to ${\cal
O}^{-4}$ which will explicitly involve 2PN order variables. But we
do not need such efforts for our background subtraction process
because all the ${\cal O}^{-4}$ order terms involve PN order
variables, which do not affect the Friedmann background. To get the
Friedmann background we set $U$, $\Phi$, $P_i$, $V$, $v_i$, $Q_i$,
and $\Pi_{ij}$ equal to zero, and set $\varrho = \varrho_b$, $\Pi =
\Pi_b$, and $p = p_b$. Then, Eqs.\ (\ref{R_00-eq-raw}),
(\ref{R_ij-eq-raw}) give \bea
   & &
       {3 \over c^2} \left\{
       {\ddot a \over a}
       + {4 \pi G \over 3} \left[ \varrho_b \left(
       1 + {\Pi_b \over c^2} \right)
       + {3 p_b \over c^2} \right]
       - {\Lambda c^2 \over 3} \right\} = 0,
   \\
   & &
       {1 \over c^2} \left\{
       {\ddot a \over a}
       + 2 {\dot a^2 \over a^2}
       - 4 \pi G \left[ \varrho_b \left( 1 + {\Pi_b \over c^2}
       \right)
       - {p_b \over c^2} \right]
       + {2 K c^2 \over a^2}
       - \Lambda c^2
       \right\} \delta^i_j = 0.
\eea Thus, we have \bea
   & & {\ddot a \over a}
       = - {4 \pi G \over 3} \left[ \varrho_b \left(
       1 + {\Pi_b \over c^2} \right)
       + {3 p_b \over c^2} \right]
       + {\Lambda c^2 \over 3},
   \label{BG-eq2} \\
   & & {\dot a^2 \over a^2}
       = {8 \pi G \over 3} \varrho_b \left( 1 + {\Pi_b \over c^2}
       \right)
       - {K c^2 \over a^2}
       + {\Lambda c^2 \over 3},
   \label{BG-eq3}
\eea which are the Friedmann equations.

By {\it subtracting} the background equations, Eqs.\
(\ref{R_00-eq-raw})-(\ref{R_ij-eq-raw}) give \bea
   & &
       {1 \over c^2} \left[ {\Delta \over a^2} U
       + 4 \pi G \left( \varrho - \varrho_b \right) \right]
       + {1 \over c^4} \Bigg\{
       3 \ddot V
       + 3 {\dot a \over a} \left( \dot U + 2 \dot V \right)
       + 6 {\ddot a \over a} U
       - {1 \over a^2} \left[ U^{,i} \left( U - V \right)_{,i}
       + 2 V \Delta U
       - 2 \Delta \Phi
       - \left( a P^i_{\;\;|i} \right)^\cdot \right]
   \nonumber \\
   & & \qquad
       + 8 \pi G \left[ \varrho v^2
       + {1 \over 2} \left( \varrho \Pi - \varrho_b \Pi_b \right)
       + {3 \over 2} \left( p - p_b \right) \right]
       \Bigg\} = 0,
   \label{R_00-eq2} \\
   & & {1 \over c^3} \left[ 2 \left( \dot V + {\dot a \over a} U
       \right)_{,i}
       + {1 \over 2 a} \left( P^j_{\;\;|ji} - \Delta P_i
       - 2 K P_i \right)
       - 8 \pi G \varrho a v_i
       \right] = 0,
   \label{R_0i-eq} \\
   & & {1 \over c^2} \left\{
       - \left[ {\Delta + 4 K \over a^2} V
       + 4 \pi G \left( \varrho - \varrho_b \right) \right]
       \delta^i_j
       + {1 \over a^2} \left( U - V \right)^{|i}_{\;\;\; j}
       \right\} = 0.
   \label{R_ij-eq}
\eea To ${\cal O}^{-2}$ order, Eq.\ (\ref{R_00-eq2}) gives \bea
   & & {\Delta \over a^2} U
       = - 4 \pi G \left( \varrho - \varrho_b \right),
   \label{Poisson-eq}
\eea which is Poisson's equation in Newton's gravity. Notice that in
the Newtonian limit of our PN approximation the homogeneous part of
density distribution is subtracted. This revised form of Poisson's
equation is consistent with Newton's gravity, and in fact, is an
improved form avoiding Jeans' swindle \cite{Lemons-1988}. The
decomposition of Eq.\ (\ref{R_ij-eq}) into trace and tracefree parts
gives \bea
   & & {\Delta + 3 K \over a^2} V = - 4 \pi G \left( \varrho - \varrho_b \right),
   \label{V-eq} \\
   & & \left( U - V \right)^{|i}_{\;\;\; j}
       = K V \delta^i_j.
   \label{UV}
\eea Thus, compared with Eq.\ (\ref{Poisson-eq}), for $K = 0$ we
have \bea
   & & V = U,
   \label{U=V}
\eea where we ignored any surface term $S$ with $\Delta S \equiv 0$
and $S_{,i|j} \equiv 0$. Notice that we have $V = U$ even in the
presence of anisotropic stress; this differs from the situation in
the perturbation approach where $U$ and $V$ in the zero-shear gauge
(setting $P^i_{\;\; |i} = 0$, see later) are different in the
presence of the anisotropic stress \cite{Bardeen-1980}. {}For later
use we take a divergence of Eq.\ (\ref{R_0i-eq}) which gives \bea
   & & {\Delta \over a^2} \left( \dot V + {\dot a \over a} U \right)
       = 4 \pi G {1 \over a} \left( \varrho v^i \right)_{|i}
       + {K \over a^3} P^i_{\;\;|i}.
   \label{div-R_0i}
\eea

Here, we reached a point to address the issue related to the
background curvature in our PN approach. By taking a divergence of
Eq.\ (\ref{R_0i-eq}) and using Eqs.\ (\ref{Poisson-eq}),
(\ref{V-eq}), and (\ref{Newtonian-mass-conservation}) we have \bea
   & &
       {1 \over c^3} \left( - 6 K \dot V
       - {2 K \over a} P^i_{\;\;|i} \right) = 0,
   \label{K-issue}
\eea which is apparently inconsistent. This is because the $K$ term
in Eq.\ (\ref{BG-eq3}) is related to the $c^{-2}$ higher order terms
in the background Friedmann equation. We can check this point by
expanding the equation to 2PN order; in that order we can find that
the above term can be removed by subtracting the background
equation. Thus, it looks like our PN approximation is properly
applicable only for $K = 0$. In this sense, our $K$ terms do not
have significance because these are related to ${\cal O}^{-2}$ order
higher terms through the background equation. At the moment, it is
unclear whether this limitation of our 1PN approach to a nearly flat
background is due to our subtraction process of the background
equations (which spread over different PN orders), or intrinsic to
the whole PN approximation. Meanwhile, recent observations favour
the flat background world model with non-vanishing cosmological
constant. We include the cosmological constant in our PN
approximation which appears only in the background Friedmann
equations. Ignoring $K$ term we simply have $V = U$.

\subsection{ADM approach}
                                              \label{sec:ADM-eqs}

The ADM equations are \cite{ADM,Bardeen-1980,NL} \bea
   & & R^{(h)} = \bar K^{ij} \bar K_{ij}
       - {2 \over 3} \bar K^2 + {16 \pi G \over c^4} E + 2 \Lambda,
   \label{E-constraint} \\
   & & \bar K^j_{i:j} - {2 \over 3} \bar K_{,i}
       = {8 \pi G \over c^4} J_i,
   \label{Mom-constraint} \\
   & & \bar K_{,0} N^{-1} - \bar K_{,i} N^i N^{-1}
       + N^{:i}_{\;\;\;i} N^{-1}
       - \bar K^{ij} \bar K_{ij}
       - {1\over 3} K^2 - {4 \pi G \over c^4} \left( E + S \right) + \Lambda =
       0,
   \label{Trace-prop} \\
   & & \bar K^i_{j,0} N^{-1}
       - \bar K^i_{j:k} N^k N^{-1}
       + \bar K_{jk} N^{i:k} N^{-1}
       - \bar K^i_k N^k_{\;\;:j} N^{-1}
   \nonumber \\
   & & \qquad = \bar K \bar K^i_j
       - \left( N^{:i}_{\;\;\;j}
       - {1\over 3} \delta^i_j N^{:k}_{\;\;\;k} \right)
       N^{-1}
       + \bar R^{(h)i}_{\;\;\;\;\;j} - {8 \pi G \over c^4} \bar
       S^i_j,
   \label{Tracefree-prop} \\
   & & E_{,0} N^{-1} - E_{,i} N^i N^{-1}
       - \bar K \left( E + {1 \over 3} S \right)
       - \bar S^{ij} \bar K_{ij}
       + N^{-2} \left( N^2 J^i \right)_{:i}
       = 0,
   \label{E-conservation} \\
   & & J_{i,0} N^{-1} - J_{i:j} N^j N^{-1}
       - J_j N^j_{\;\;:i} N^{-1} - \bar K J_i
       + E N_{,i} N^{-1}
       + S^j_{i:j}
       + S_i^j N_{,j} N^{-1} = 0.
   \label{Mom-conservation}
\eea Using the ADM quantities presented in Sec.\
\ref{sec:ADM-quantities} we can show that Eqs.\
(\ref{E-constraint})-(\ref{Mom-conservation}) give the same
equations we already have derived from Einstein's equations and the
energy and momentum conservations. Equation (\ref{E-constraint})
gives Eqs.\ (\ref{BG-eq3}), (\ref{V-eq}). Equation
(\ref{Mom-constraint}) gives Eq.\ (\ref{R_0i-eq}). Equation
(\ref{Trace-prop}) gives Eq.\ (\ref{R_00-eq-raw}). Equation
(\ref{Tracefree-prop}) gives the tracefree part of Eq.\
(\ref{R_ij-eq}). Equation (\ref{E-conservation}),
(\ref{Mom-conservation}) give Eqs.\ (\ref{E-conserv-0}),
(\ref{Mom-conserv-0}), respectively. The ADM equations are often
used in the PN approach
\cite{Blanchet-etal-1990,Shibata-Asada-1995,Asada-Futamase-1997,Takada-Futamase-1999}.
We present the ADM approach because the ADM equations show the fully
nonlinear structure of Einstein's gravity in a form suitable for
numerical treatment. In fact, Eq.\ (\ref{v^*_i-eq}) can be derived
from Eq.\ (\ref{Mom-conservation}) using the identification made in
Eq.\ (\ref{J_i-v^*}).

%
%
\section{1PN Equations}
                                            \label{sec:Equations}

\subsection{Complete equations to 1PN order}

Here we summarize the complete set of equations valid to 1PN order
in the cosmological situation. The background variables $a$,
$\varrho_b$, $\Pi_b$ and $p_b$ are provided by solving Eqs.\
(\ref{BG-eq1}), (\ref{BG-eq1-1}), (\ref{BG-eq2}), and (\ref{BG-eq3})
\bea
   & & {\ddot a \over a}
       = - {4 \pi G \over 3} \left[ \varrho_b \left(
       1 + {\Pi_b \over c^2} \right)
       + {3 p_b \over c^2} \right]
       + {\Lambda c^2 \over 3},
   \label{BG-eq2-c} \\
   & & {\dot a^2 \over a^2}
       = {8 \pi G \over 3} \varrho_b \left( 1 + {\Pi_b \over c^2}
       \right)
       - {K c^2 \over a^2}
       + {\Lambda c^2 \over 3},
   \label{BG-eq3-c} \\
   & &
       \dot \mu_b + 3 {\dot a \over a}
       \left( \mu_b + p_b \right) = 0,
   \label{BG-eq1-c} \\
   & &
       \dot \varrho_b + 3 {\dot a \over a} \varrho_b = 0,
   \label{BG-eq1-1-c}
\eea where $\mu_b \equiv \varrho_b c^2 \left( 1 + \Pi_b/c^2
\right)$. In Eqs.\ (\ref{BG-eq2-c})-(\ref{BG-eq1-c}) only two are
independent, and from Eqs.\ (\ref{BG-eq1-c}), (\ref{BG-eq1-1-c}) we
have $\varrho_b \dot \Pi_b + 3 (\dot a /a) p_b = 0$.

To the Newtonian order we have Eqs.\
(\ref{Newtonian-mass-conservation}),
(\ref{Newtonian-momentum-conservation}), and (\ref{Poisson-eq}) \bea
   & &
       {1 \over a^3} \left( a^3 \varrho \right)^\cdot
       + {1 \over a} \nabla_i \left( \varrho v^i \right) = 0,
   \label{Newtonian-mass-conservation-c} \\
   & &
       {1 \over a} \left( a v_i \right)^\cdot
       + {1 \over a} v^j \nabla_j v_i
       + {1 \over a \varrho} \left( \nabla_i p
       + \nabla_j \Pi^j_i \right)
       - {1 \over a} \nabla_i U = 0,
   \label{Newtonian-momentum-conservation-c} \\
   & & {\Delta \over a^2} U
       + 4 \pi G \left( \varrho - \varrho_b \right) = 0.
   \label{Poisson-eq-c}
\eea In the Newtonian case energy conservation and mass conservation
provide an additional equation which is Eq.\ (\ref{adiabatic-case})
\bea
   & & \left( {\partial \over \partial t}
       + {1 \over a} {\bf v} \cdot \nabla \right) \Pi
       + \left( 3 {\dot a \over a}
       + {1 \over a} \nabla \cdot {\bf v} \right)
       {p \over \varrho}
       + {1 \over \varrho a} \left( Q^i_{\;\;|i}
       + \Pi^i_j v^j_{\;\;|i} \right)
       = 0.
   \label{adiabatic-eq-c}
\eea In the non-expanding background this equation gives the well
known energy conservation equation in the Newtonian theory, see for
example Eq.\ (2.36) in \cite{Shu-1992}. The Newtonian order
hydrodynamic equations are valid in the presence of general
background curvature $K$.

In the following we set $K = 0$ to the 1PN  order, because our PN
expansion fails to apply in the presence of $K$ term. Thus we have
$V = U$. The energy and the momentum conservation equations valid to
1PN order are derived in Eqs.\ (\ref{E-conserv-pert-0}),
(\ref{Mom-conserv-0}), and these are \bea
   & &
       {1 \over a^3} \left( a^3 \sigma \right)^\cdot
       + {1 \over a} \left[ \sigma v^i
       + {1 \over c^2} \left( Q^i
       + \Pi^i_j v^j \right)
       \right]_{|i}
       + {1 \over c^2} \varrho \left( \dot U + {\dot a \over a} v^2
       - {\dot p \over \varrho} \right)
       = 0,
   \label{E-conserv-0-c} \\
   & & {1 \over a^4} \left\{ a^4 \left[ \sigma v_i
       + {1 \over c^2} \left( Q_i + \Pi_{ij} v^j \right)
       \right] \right\}^\cdot
       + {1 \over a} \left[ \sigma v_i v^j
       + \Pi^j_i
       + {1 \over c^2} \left(
       Q^j v_i + Q_i v^j
       - 4 U \Pi^j_i \right)
       \right]_{|j}
   \nonumber \\
   & & \qquad
       + {1 \over a} \left( 1 - {1 \over c^2} 2 U \right) p_{,i}
       - {1 \over a} \left( \sigma
       + {1 \over c^2} \varrho v^2 \right) U_{,i}
   \nonumber \\
   & & \qquad
       + {1 \over c^2} \varrho \left[
       4 v_i \left( {\partial \over \partial t}
       + {1 \over a} {\bf v} \cdot \nabla \right) U
       + {2 \over a} \left( U^2 - \Phi \right)_{,i}
       - {1 \over a} \left( a P_i \right)^\cdot
       - {2 \over a} v^j P_{[i|j]}
       + {4 \over \varrho a} U_{,j} \Pi^j_i
       \right]
       = 0,
   \label{Mom-conserv-0-c}
\eea where \bea
   & & \sigma \equiv \varrho \left[ 1
       + {1 \over c^2} \left( v^2 + 2 U + \Pi + {p \over
       \varrho} \right)
       \right].
   \label{sigma-def-c}
\eea To the Newtonian order Eqs.\ (\ref{E-conserv-0-c}),
(\ref{Mom-conserv-0-c}) for $\varrho$ and $v_i$ reduce to Eqs.\
(\ref{Newtonian-mass-conservation-c}),
(\ref{Newtonian-momentum-conservation-c}), respectively. From Eq.\
(\ref{E-conserv-00}), (\ref{v^*_i-eq}) we can derive alternative
forms
 \bea
   & & {1 \over a^3} \left( a^3 \varrho^* \right)^\cdot
       + {1 \over a} \left( \varrho^* v^i \right)_{|i}
       = - {1 \over c^2} \left[
       {1 \over a} \left( Q^i_{\;\;|i}
       + \Pi^i_j v^j_{\;\;|i} \right)
       + \varrho \left( {\partial \over \partial t}
       + {1 \over a} {\bf v} \cdot \nabla \right) \Pi
       + \left( 3 {\dot a \over a}
       + {1 \over a} \nabla \cdot {\bf v} \right) p
       \right],
   \label{E-conserv-00-c} \\
   & & {1 \over a} \left( a v^*_i \right)^\cdot
       + {1 \over a} v^*_{i|j} v^j
       = {1 \over a} \left\{
       U_{,i}
       + {1 \over c^2} \left[ \left( {3 \over 2} v^2
       - U + \Pi + 3 {p \over \varrho} \right) U_{,i}
       + 2 \Phi_{,i}
       - v^j P_{j|i} \right] \right\}
   \nonumber \\
   & & \qquad
       - {1 \over \varrho^* a}
       \left\{ \left( 1 + {1 \over c^2} 2 U \right)
       \left[ p \delta^j_i
       + \Pi^j_i
       + {1 \over c^2} \left( Q^j v_i
       - 2 V \Pi^j_i
       - \Pi_{ik} v^j v^k \right) \right] \right\}_{|j}
   \nonumber \\
   & & \qquad
       + {1 \over c^2} {v^*_i \over \varrho^*}
       \left[ {1 \over a} \left( Q^j + \Pi^j_k v^k \right)_{|j}
       + \varrho \left( {\partial \over \partial t}
       + {1 \over a} {\bf v} \cdot \nabla \right) \Pi
       + \left( 3 {\dot a \over a}
       + {1 \over a} \nabla \cdot {\bf v} \right) p
       \right],
   \label{v^*_i-eq-c}
\eea where \bea
   & & \varrho^*
       \equiv \varrho \left[ 1 + {1 \over c^2}
       \left( {1 \over 2} v^2 + 3 U \right) \right],
   \nonumber \\
   & & v^*_i \equiv v_i
       + {1 \over c^2} \left[ \left( {1 \over 2} v^2 + 3 U +
       \Pi + {p \over \varrho} \right) v_i
       - P_i
       + {1 \over \varrho} \left( Q_i
       + \Pi_{ij} v^j \right) \right].
   \label{varrho*-def}
\eea The hydrodynamic and thermodynamic variables $\Pi$, $p$, $Q_i$
and $\Pi_{ij}$ should be provided by specifying the equations of
state and the thermodynamic state of the system under consideration.

The metric variables $U$, $\Phi$ and $P_i$ can be expressed in terms
of the Newtonian variables $\varrho$, $v_i$, $\Pi$, $p$, $Q_i$ and
$\Pi_{ij}$ by using Einstein's equations. Equations
(\ref{Poisson-eq}), (\ref{V-eq}) give \bea
   & & {\Delta \over a^2} U
       = - 4 \pi G \left( \varrho - \varrho_b \right),
   \label{Pert1}
\eea which already determines $U$. Equations (\ref{R_00-eq2}),
(\ref{R_0i-eq}), and (\ref{div-R_0i}) give \bea
   & & {\Delta \over a^2} U
       + 4 \pi G \left( \varrho - \varrho_b \right)
       + {1 \over c^2} \Bigg\{
       {1 \over a^2} \left[
       2 \Delta \Phi
       - 2 U \Delta U
       + \left( a P^i_{\;\;|i} \right)^\cdot \right]
       + 3 \ddot U
       + 9 {\dot a \over a} \dot U
       + 6 {\ddot a \over a} U
   \nonumber \\
   & & \qquad
       + 8 \pi G \left[ \varrho v^2
       + {1 \over 2} \left( \varrho \Pi - \varrho_b \Pi_b \right)
       + {3 \over 2} \left( p - p_b \right) \right]
       \Bigg\} = 0,
   \label{Pert3} \\
   & &
       {\Delta \over a^2} P_i
       = - 16 \pi G \varrho v_i
       + {1 \over a} \left( {1 \over a} P^j_{\;\;|j}
       + 4 \dot U + 4 {\dot a \over a} U \right)_{,i},
   \label{Pert4} \\
   & & {\Delta \over a^2} \dot U
       = 4 \pi G \left[ {1 \over a} \left( \varrho v^i \right)_{|i}
       + {\dot a \over a} \left( \varrho - \varrho_b \right) \right].
   \label{Pert5}
\eea Notice that to 1PN order Einstein's equations do not involve
the anisotropic stress or flux term. To 1PN order Eq.\ (\ref{Pert3})
determines $\Phi$. Our conservation equations (\ref{E-conserv-0-c}),
(\ref{Mom-conserv-0-c}) contain $\dot U$ terms. In order to handle
these terms it was suggested in \cite{Shibata-Asada-1995} that the
Poisson-type equation in Eq.\ (\ref{Pert5}) provides better
numerical accuracy. Notice that, whereas only the spatial gradient
of the potential $U$ appears in the Newtonian limit, we have bare
$U$ terms present in the 1PN order \cite{Ryu-2005}.

In order to handle these equations we have the freedom to take one
temporal gauge condition. This corresponds to imposing a condition
on $P^i_{\;\;|i}$ or $\Phi$. Gauge related issues will be addressed
in detail in Sec.\ \ref{sec:Gauges}. There, we will show that all
our variables in this section are spatially gauge-invariant, and are
temporally gauge-ready.

\subsection{Ideal fluid case}
                                                  \label{sec:IF}

We consider an ideal fluid, i.e., $Q_i \equiv 0 \equiv \Pi_{ij}$,
and assume the adiabatic condition Eq.\ (\ref{adiabatic-eq-c})
applies. Thus, the internal energy is determined by the energy
conservation equation \bea
   & & \left( {\partial \over \partial t}
       + {1 \over a} {\bf v} \cdot \nabla \right) \Pi
       + \left( 3 {\dot a \over a}
       + {1 \over a} \nabla \cdot {\bf v} \right)
       {p \over \varrho}
       = 0.
   \label{adiabatic-eq-IF}
\eea To the background order we have Eqs.\
(\ref{BG-eq2-c})-(\ref{BG-eq1-1-c}). To the Newtonian order we have
Eqs.\ (\ref{Newtonian-mass-conservation-c})-(\ref{adiabatic-eq-c}).
Equations (\ref{E-conserv-0-c}), (\ref{Mom-conserv-0-c}) become \bea
   & & {1 \over a^3}
       \left( a^3 \sigma \right)^\cdot
       + {1 \over a} \left( \sigma v^i \right)_{|i}
       + {1 \over c^2} \varrho \left( \dot U
       + {\dot a \over a} v^2
       - {\dot p \over \varrho} \right) = 0,
   \label{E-conserv-0-IF} \\
   & & {1 \over a^4} \left( a^4 \sigma v_i \right)^\cdot
       + {1 \over a} \left( \sigma v_i v^j \right)_{|j}
       - {1 \over a} \left( \sigma
       + {1 \over c^2} \varrho v^2 \right) U_{,i}
       + {1 \over a} \left( 1 - {1 \over c^2} 2 U \right) p_{,i}
   \nonumber \\
   & & \qquad
       + {1 \over c^2} \varrho \left[
       4 v_i \left( {\partial \over \partial t}
       + {1 \over a} {\bf v} \cdot \nabla \right) U
       + {2 \over a} \left( U^2 - \Phi \right)_{,i}
       - {1 \over a} \left( a P_i \right)^\cdot
       - {2 \over a} v^j P_{[i|j]}
       \right]
       = 0,
   \label{Mom-conserv-0-IF}
\eea where $\sigma$ is given in Eq.\ (\ref{sigma-def-c}). These
equations can be written as \bea
   & & {1 \over a^3} \left( a^3 \varrho \right)^\cdot
       + {1 \over a} \left( \varrho v^i \right)_{|i}
       + {1 \over c^2} \varrho \left( {\partial \over \partial t}
       + {1 \over a} {\bf v} \cdot \nabla \right)
       \left( {1 \over 2} v^2 + 3 U \right) = 0,
   \label{mass-conservation-IF} \\
   & & {1 \over a} \left( a v_i \right)^\cdot
       + {1 \over a} v_{i|j} v^j
       - {1 \over a} U_{,i}
       + {1 \over a} {p_{,i} \over \varrho}
       + {1 \over c^2} \Bigg[
       - {1 \over a} \left( v^2 + 4 U + \Pi + {p \over \varrho}
       \right) {p_{,i} \over \varrho}
   \nonumber \\
   & & \qquad
       + v_i \left( {\partial \over \partial t}
       + {1 \over a} {\bf v} \cdot \nabla \right)
       \left( {1 \over 2} v^2 + 3 U  + \Pi + {p \over \varrho} \right)
       - {1 \over a} v^2 U_{,i}
       + {2 \over a} \left( U^2 - \Phi \right)_{,i}
       - {1 \over a} \left( a P_i \right)^\cdot
       - {2 \over a} v^j P_{[i|j]}
       \Bigg]
       = 0.
   \label{momentum-conservation-IF}
\eea Alternative forms follow from Eqs.\ (\ref{E-conserv-00-c}),
(\ref{v^*_i-eq-c}) \bea
   & & {1 \over a^3} \left( a^3 \varrho^* \right)^\cdot
       + {1 \over a} \left( \varrho^* v^i \right)_{|i}
       = 0,
   \\
   & & {1 \over a} \left( a v^*_i \right)^\cdot
       + {1 \over a} v^*_{i|j} v^j
       = - {1 \over a} \left( 1 + {1 \over c^2} 2 U \right)
       {p_{,i} \over \varrho^*}
       + {1 \over a}
       \left[ 1 + {1 \over c^2} \left( {3 \over 2} v^2
       - U + \Pi + {p \over \varrho} \right) \right] U_{,i}
       + {1 \over c^2} {1 \over a} \left( 2 \Phi_{,i}
       - v^j P_{j|i} \right),
   \label{v^*_i-eq-IF}
\eea where \bea
   & & \varrho^* \equiv \varrho \left[ 1 + {1 \over c^2} \left(
       {1 \over 2} v^2 + 3 U \right) \right], \quad
       v^*_i
       \equiv v_i
       + {1 \over c^2} \left[ \left( {1 \over 2} v^2 + 3 U +
       \Pi + {p \over \varrho} \right) v_i
       - P_i \right].
\eea All these equivalent three sets of equations are written in
Eulerian forms. Einstein's equations in (\ref{Pert3})-(\ref{Pert5})
will provide the metric perturbation variables $U$, $\Phi$ and $P_i$
in terms of the Newtonian fluid variables ($\varrho$, $v^i$, $\Pi$,
and $p$), after taking one temporal gauge condition. The pressure
term should be provided by an equation of state. In the case of a
zero-pressure fluid with vanishing internal energy we can set $p = 0
= \Pi$. We can take any set of equations of motion depending on the
mathematical convenience in numerical treatments.

%
%
\section{Gauge issue}
                                                 \label{sec:Gauges}

\subsection{Gauge transformation}

We consider the following transformation between two coordinates
$x^a$ and $\hat x^a$ \bea
   & & \hat  x^a
       \equiv x^a + \tilde \xi^a (x^e).
   \label{GT-x}
\eea {}For a tensor quantity we use the tensor transformation
property between $x^a$ and $\hat x^a$ coordinates \bea
   & & \tilde t_{ab} (x^e)
       = {\partial \hat x^c \over \partial x^a}
       {\partial \hat x^d \over \partial x^b}
       \hat {\tilde t}_{cd} (\hat x^e).
   \label{coord-tr}
\eea Comparing tensor quantities at the same spacetime point, $x^a$,
we can derive the gauge transformation property of a tensor
quantity, see Eq.\ (226) in \cite{NL}, \bea
   {\hat {\tilde t}}_{ab} (x^e)
   &=& \tilde t_{ab} (x^e) - 2 \tilde t_{c(a} \tilde \xi^c_{\;\;,b)}
       - \tilde t_{ab,c} \tilde \xi^c
       + 2 \tilde t_{c(a} \tilde \xi^d_{\;\;,b)} \tilde \xi^c_{\;\;,d}
       + \tilde t_{cd} \tilde \xi^c_{\;\;,a} \tilde \xi^d_{\;\;,b}
   \nonumber \\
   & &
       + \tilde \xi^d \left(
       2 \tilde \xi^c_{\;\;,(a} \tilde t_{b)c,d}
       + 2 \tilde t_{c(a} \tilde \xi^c_{\;\;,b)d}
       + {1 \over 2} \tilde t_{ab,cd} \tilde \xi^c
       + \tilde t_{ab,c} \tilde \xi^c_{\;\;,d}
       \right).
   \label{GT-tensor}
\eea We considered $\tilde \xi^a$ to the second perturbational order
which will turn out to be sufficient for the 1PN order. In the
following we will consider the gauge transformation properties of
the metric and the energy-momentum variables.

As the metric we consider the following more generalized form \bea
   & & \tilde g_{00} \equiv - \left[ 1 - {1 \over c^2} 2 U
       + {1 \over c^4} \left( 2 U^2 - 4 \Phi \right) \right]
       + {\cal O}^{-6},
   \nonumber \\
   & & \tilde g_{0i} \equiv - {1 \over c^3} a P_i + {\cal O}^{-5},
   \nonumber \\
   & & \tilde g_{ij} \equiv a^2 \left[
       \left( 1 + {1 \over c^2} 2 V \right) \gamma_{ij}
       + {1 \over c^2} \left( 2 C_{,i|j}
       + 2 C_{(i|j)} + 2 C_{ij} \right)
       \right]
       + {\cal O}^{-4},
   \label{metric-general}
\eea where $C_i$ is transverse ($C^i_{\; |i} \equiv 0$), and
$C_{ij}$ is transverse and tracefree ($C^j_{i|j} \equiv 0 \equiv
C^i_i$); indices of $C_i$ and $C_{ij}$ are based on $\gamma_{ij}$.
In Eq.\ (\ref{metric-general}) we introduced 10 independent metric
components: $U$ and $\Phi$ together (1-component), $P_i$
(3-components), $V$ (1-component), $C$ (1-component), $C_i$
(2-components), and $C_{ij}$ (2-components). It is known that
gravitational waves show up in the 2.5PN order
\cite{Chandrasekhar-Esposito-1970}. Thus, we ignore the transverse
and tracefree part, i.e., set $C_{ij} \equiv 0$ to 1PN order.

We wish to keep the metric in the form of Eq.\
(\ref{metric-general}) in {\it any} coordinate system. Thus, we take
the transformation variable $\tilde \xi^a$ to be a PN-order
quantity. We consider coordinate transformations which satisfy \bea
   & & \tilde \xi^0
       \equiv {1 \over c} \xi^{(2)0}
       + {1 \over c^3} \xi^{(4)0} + \dots,
       \quad
       \tilde \xi^i \equiv {1 \over c^2} {1 \over a} \xi^{(2)i} + \dots,
\eea where the indices of the $\xi^{(2)i}$'s are based on
$\gamma_{ij}$.

The gauge transformation of Eq.\ (\ref{GT-tensor}) applied to the
metric gives \bea
   \hat U
   &=& U + \dot \xi^{(2)0},
   \label{metric-GT-1} \\
   \hat \Phi
   &=& \Phi + {1 \over 2} \dot \xi^{(4)0}
       - {1 \over 2} \dot U \xi^{(2)0}
       - {1 \over 2 a} U_{,i} \xi^{(2)i}
       - {1 \over 2} \xi^{(2)0} \ddot \xi^{(2)0}
       - {1 \over 4} \dot \xi^{(2)0} \dot \xi^{(2)0}
       - {1 \over 2} \left( {1 \over a}
       \xi^{(2)i} \xi^{(2)0}_{\;\;\;\;\;\;,i}
       \right)^\cdot,
   \label{metric-GT-2} \\
   \xi^{(2)0}_{\;\;\;\;\;\;\;,i}
   &=& 0,
   \label{metric-GT-3} \\
   \hat P_i
   &=& P_i
       - {1 \over a} \xi^{(4)0}_{\;\;\;\;\;\;,i}
       + a \left( {1 \over a} \xi^{(2)}_i \right)^\cdot
       + {2 \over a} \left( U + \dot \xi^{(2)0} \right)
       \xi^{(2)0}_{\;\;\;\;\;\;,i}
       + {1 \over a} \dot \xi^{(2)0}_{\;\;\;\;\;\;,i} \xi^{(2)0}
       + {1 \over a^2}
       \left( \xi^{(2)j} \xi^{(2)0}_{\;\;\;\;\;\;,j} \right)_{,i},
   \label{metric-GT-4} \\
   \left( \Delta + 3 K \right) \Delta \hat C
   &=& \left( \Delta + 3 K \right) \left( \Delta C
       - {1 \over a} \xi^{(2)i}_{\;\;\;\;\;\;|i} \right)
       + {\Delta \over 4 a^2} \left( \xi^{(2)0,i}
       \xi^{(2)0}_{\;\;\;\;\;\;,i} \right)
       - {3 \over 4 a^2} \left( \xi^{(2)0}_{\;\;\;\;\;\;,i}
       \xi^{(2)0}_{\;\;\;\;\;\;,j} \right)^{|ij},
   \label{metric-GT-5} \\
   \left( \Delta + 2 K \right) \hat C_i
   &=& \left( \Delta + 2 K \right) \left( C_i
       - {1 \over a} \xi^{(2)}_i \right)
       - {1 \over 3 a} \xi^{(2)j}_{\;\;\;\;\;\;|ji}
       - {1 \over a^2} \left( \xi^{(2)0}_{\;\;\;\;\;\;,i}
       \xi^{(2)0}_{\;\;\;\;\;\;,j} \right)^{|j}
   \nonumber \\
   & &
       + {1 \over a} \nabla_i \Delta^{-1} \left\{
       {4 \over 3} \left( \Delta + 3 K \right)
       \xi^{(2)j}_{\;\;\;\;\;\;|j}
       + {1 \over a}
       \left[ \left( \xi^{(2)0}_{\;\;\;\;\;\;,k}
       \xi^{(2)0}_{\;\;\;\;\;\;,j} \right)^{|kj} \right]
       \right\},
   \label{metric-GT-6} \\
   \left( \Delta + 3 K \right) \hat V
   &=& \left( \Delta + 3 K \right) \left( V
       - {\dot a \over a} \xi^{(2)0} \right)
       - {\Delta + 2 K \over 4 a^2} \left( \xi^{(2)0,i}
       \xi^{(2)0}_{\;\;\;\;\;\;,i} \right)
       + {1 \over 4 a^2}
       \left( \xi^{(2)0}_{\;\;\;\;\;\;,i}
       \xi^{(2)0}_{\;\;\;\;\;\;,j} \right)^{|ij}.
   \label{metric-GT-7}
\eea Equations (\ref{metric-GT-1}), (\ref{metric-GT-2}) follow from
the transformation of $\tilde g_{00}$; Eqs.\ (\ref{metric-GT-3}),
(\ref{metric-GT-4}) follow from $\tilde g_{0i}$; Eqs.\
(\ref{metric-GT-5})-(\ref{metric-GT-7}) follow from $\tilde g_{ij}$.

Equation (\ref{metric-GT-3}) shows that $\xi^{(2)0}$ is spatially
constant, \bea
   & & \xi^{(2)0} = \xi^{(2)0} (t).
   \label{xi-0}
\eea After this, Eq.\ (\ref{metric-GT-5}) gives \bea
   & & \Delta \hat C
       = \Delta C
       - {1 \over a} \xi^{(2)i}_{\;\;\;\;\;\;|i}.
\eea Thus, by choosing $C \equiv 0$ ($\Delta C \equiv 0$ is enough)
as the gauge condition (this implies that we set $C = 0$ to be valid
in any coordinate), we have \bea
   & & \xi^{(2)i}_{\;\;\;\;\;\;|i} = 0.
   \label{xi-ii}
\eea Imposing the conditions in Eqs.\ (\ref{xi-0}), (\ref{xi-ii}),
Eq.\ (\ref{metric-GT-6}) gives \bea
   & & \hat C_i
       = C_i - {1 \over a} \xi^{(2)}_i.
\eea Thus, by choosing $C_i \equiv 0$ to be valid in any coordinate,
i.e., choosing $C_i \equiv 0$ as the gauge condition, we have \bea
   & & \xi^{(2)}_i = 0.
   \label{xi-i}
\eea Under the conditions in Eqs.\ (\ref{xi-0}), (\ref{xi-i}) the
remaining metric perturbation variables transform as \bea
   & & \hat U = U + \dot \xi^{(2)0}, \quad
       \hat V = V - {\dot a \over a} \xi^{(2)0}, \quad
       \hat \Phi = \Phi + {1 \over 2} \dot \xi^{(4)0}
       - {1 \over 2} \dot U \xi^{(2)0}
       - {1 \over 2} \xi^{(2)0} \ddot \xi^{(2)0}
       - {1 \over 4} \dot \xi^{(2)0} \dot \xi^{(2)0},
   \nonumber \\
   & &
       \hat P_i = P_i - {1 \over a} \xi^{(4)0}_{\;\;\;\;\;\;,i}.
\eea Notice that for non-vanishing $\xi^{(2)0}$, $U$ and $V$
transform differently under the gauge transformation even in the
flat cosmological background. Since the spatially constant
$\xi^{(2)0}$ can be absorbed by a global redefinition of the time
coordinate, without losing generality we can set $\xi^{(2)0}$ equal
to zero \bea
   & & \xi^{(2)0} \equiv 0,
   \label{xi-0=0}
\eea thus allowing $V = U$ in any coordinate. If we set $\xi^{(2)0}
= 0$ but do not take the spatial gauge which fixes $\xi^{(2)}_i$,
from Eqs.\ (\ref{metric-GT-1})-(\ref{metric-GT-7}), the metric
variables transform as \bea
   \hat U
   &=& U,
   \label{metric-GT2-1} \\
   \hat \Phi
   &=& \Phi + {1 \over 2} \dot \xi^{(4)0}
       - {1 \over 2 a} U_{,i} \xi^{(2)i},
   \label{metric-GT2-2} \\
   \hat P_i
   &=& P_i
       - {1 \over a} \xi^{(4)0}_{\;\;\;\;\;\;,i}
       + a \left( {1 \over a} \xi^{(2)}_i \right)^\cdot,
   \label{metric-GT2-4} \\
   \Delta \hat C
   &=& \Delta C
       - {1 \over a} \xi^{(2)i}_{\;\;\;\;\;\;|i},
   \label{metric-GT2-5} \\
   \left( \Delta + 2 K \right) \hat C_i
   &=& \left( \Delta + 2 K \right) \left( C_i
       - {1 \over a} \xi^{(2)}_i \right)
       - {1 \over 3 a} \xi^{(2)j}_{\;\;\;\;\;\;|ji}
       + {4 \over 3 a} \nabla_i \Delta^{-1} \left(
       \Delta + 3 K \right)
       \xi^{(2)j}_{\;\;\;\;\;\;|j},
   \label{metric-GT2-6} \\
   \hat V
   &=& V.
   \label{metric-GT2-7}
\eea

Taking the spatial gauge conditions $C = 0 = C_i$ we have
$\xi^{(2)}_i = 0 = \xi^{(2)0}$, and the remaining metric
perturbation variables transform as \bea
   & & \hat U = U, \quad
       \hat V = V, \quad
       \hat \Phi = \Phi + {1 \over 2} \dot \xi^{(4)0}, \quad
       \hat P_i = P_i - {1 \over a} \xi^{(4)0}_{\;\;\;\;\;\;,i}.
\eea In the perturbation analysis we call $C \equiv 0 \equiv C_i$
the spatial $C$-gauge \cite{NL}. Apparently, under these gauge
conditions the spatial gauge transformation function to 1PN order,
$\xi^{(2)}_i$, is fixed completely, i.e., $\xi^{(2)}_i = 0$. By
taking such gauge conditions, the only remaining gauge
transformation function to 1PN order is $\xi^{(4)0}$. This temporal
gauge transformation function affects only $\Phi$ and $P_i$. If we
take $\Phi \equiv 0$ as the temporal gauge condition, we have $\dot
\xi^{(4)0} = 0$, thus $\xi^{(4)0}$ does not vanish even after
imposing the temporal gauge condition and has general dependence on
spatial coordinate $\xi^{(4)0} ({\bf x})$. Thus, in this gauge, even
after imposing the temporal gauge condition the temporal gauge mode
is not fixed completely. Whereas, if we take $P^i_{\;\;|i} \equiv 0$
as the temporal gauge condition, we have $\xi^{(4)0} = 0$, thus the
temporal gauge condition is fixed completely. In fact, from the
gauge transformation properties of $\Phi$ and $P_i$ we can make the
following combinations \bea
   & & \Phi_{,i} + {1 \over 2} \left( a P_i \right)^\cdot, \quad
       \Delta \Phi + {1 \over 2} \left( a P^i_{\;\;|i}
       \right)^\cdot,
   \label{GI-temporal}
\eea which are invariant under the temporal gauge transformation,
i.e., temporally gauge-invariant. Meanwhile, $U$ and $V$ are already
temporally gauge-invariant. Since our spatial $C$-gauge also has
removed the spatial gauge transformation function completely, we can
correspond each remaining variable to a unique (spatially and
temporally) gauge-invariant combination. Thus, in this sense, we can
equivalently regard all our remaining variables as gauge-invariant
ones. {}For example, for $K = 0$, from Eqs.\
(\ref{metric-GT2-1})-(\ref{metric-GT2-7}) we can show \bea
   & & \hat P_i + a \left( \hat C_i + \hat C_{,i} \right)^\cdot
       = P_i + a \left( C_i + C_{,i} \right)^\cdot
       - {1 \over a} \xi^{(4)0}_{\;\;\;\;\;\;,i}.
   \label{GI-spatial}
\eea Thus, $P_i + a \left( C_i + C_{,i} \right)^\cdot$ is spatially
gauge-invariant and becomes $P_i$ under the spatial $C$-gauge. This
implies that $P_i$ under the spatial $C$-gauge is equivalent to a
unique gauge-invariant combination $P_i + a \left( C_i + C_{,i}
\right)^\cdot$. Similarly, in the case of the temporal gauge, from
Eq.\ (\ref{GI-temporal}) we can show that $\Delta \Phi$ under the
$P^i_{\;\;|i} = 0$ gauge is the same as a unique gauge-invariant
combination $\Delta \Phi + {1 \over 2} ( a P^i_{\;\;|i} )^\cdot$. In
this sense, under gauge conditions which fix the gauge mode
completely, the remaining variables can be regarded as equivalently
gauge-invariant ones with corresponding gauge-invariant variables.

In Eq.\ (\ref{metric}) we began our 1PN analysis by choosing the
spatial $C$-gauge \bea
   & & C \equiv 0 \equiv C_i.
   \label{C-gauge}
\eea By examining Eqs.\ (\ref{metric-GT2-1})-(\ref{metric-GT2-7}) we
notice that the spatial $C$-gauge is most economic in fixing the
spatial gauge mode completely without due alternative. In this sense
the spatial $C$-gauge can be regarded as a unique choice and we do
not lose any mathematical convenience by taking this spatial gauge
condition. In the literature these conditions are often expressed in
the following forms. The spatial component of the harmonic gauge
condition sets the 1PN part of \bea
   & & \tilde g^{ab} \tilde \Gamma^i_{ab}
       = - {1 \over \sqrt{-\tilde g}} \left( \sqrt{-\tilde g} \tilde
       g^{ic} \right)_{,c}
       = {1 \over c^2} {\Delta \over a^2} \left( C^{,i} + C^i
       \right) + L^{-1} {\cal O}^{-4},
   \label{spatial-harmonic}
\eea equal to zero. We can also set the 1PN part of \bea
   & & \tilde g^{jk} \left( \tilde g_{ij,k}
       - {1 \over 3} \tilde g_{jk,i} \right)
       = {1 \over c^2} \Delta \left( {4 \over 3} C_{,i} + C_i
       \right) + L^{-1} {\cal O}^{-4},
   \label{spatial-harmonic-2}
\eea equal to zero; notice that in Eqs.\ (\ref{spatial-harmonic}),
(\ref{spatial-harmonic-2}) we assumed $K = 0$. In either case we
arrive at Eq.\ (\ref{C-gauge}). We have shown that under the
conditions in Eq.\ (\ref{C-gauge}), the spatial gauge transformation
is fixed completely without losing any generality or convenience.
Whereas, we have not chosen the temporal gauge condition which can
be best achieved by imposing a condition on $P^i_{\;\;|i}$. We call
this the gauge-ready strategy \cite{Hwang-1991,NL}. It is convenient
to choose this remaining temporal gauge condition depending on the
problem we encounter; or to try many different ones in order to find
out the best suitable one or ones.

Now, let us consider the gauge transformation property of the
energy-momentum tensor. We set $\xi^{(2)0} \equiv 0$. The gauge
transformation property of Eq.\ (\ref{GT-tensor}) applied to the
energy-momentum tensor in Eq.\ (\ref{Tab}) gives \bea
   \hat \varrho
   &=& \varrho
       + {1 \over c^2} \varrho^{(2)},
   \label{EM-gauge-1} \\
   \hat \Pi
   &=& \Pi
       - {1 \over a} {\varrho_{,i} \over \varrho} \xi^{(2)i}
       - {\varrho^{(2)} \over \varrho},
   \label{EM-gauge-2} \\
   \hat v_i
   &=& v_i
       + {1 \over c^2} v^{(2)}_i,
   \label{EM-gauge-3} \\
   \hat Q_i
   &=& Q_i
       - \varrho \left[ v^{(2)}_i
       - a \left( {1 \over a} \xi^{(2)}_i \right)^\cdot
       + {1 \over a} v_j \xi^{(2)j}_{\;\;\;\;\;\;|i}
       + {1 \over a} v_{i|j} \xi^{(2)j} \right],
   \label{EM-gauge-4} \\
   \hat p
   &=& p
       - {1 \over c^2} {1 \over a} \left(
       p_{,i} \xi^{(2)i}
       + {2 \over 3} p \xi^{(2)i}_{\;\;\;\;\;\;|i}
       + {2 \over 3} \Pi^i_j \xi^{(2)j}_{\;\;\;\;\;\;|i} \right),
   \label{EM-gauge-5} \\
   \hat \Pi_{ij}
   &=& \Pi_{ij}
       - {1 \over c^2} {1 \over a} \left[
       2 p \left( \xi^{(2)}_{(i|j)}
       - {1 \over 3} \gamma_{ij} \xi^{(2)k}_{\;\;\;\;\;\;|k} \right)
       + \Pi_{ij|k} \xi^{(2)k}
       + 2 \Pi_{k(i} \xi^{(2)k}_{\;\;\;\;\;\; |j)}
       - {2 \over 3} \gamma_{ij} \Pi^k_l \xi^{(2)l}_{\;\;\;\;\;\;|k}
       \right].
   \label{EM-gauge-6}
\eea Equations (\ref{EM-gauge-1}), (\ref{EM-gauge-2}) follow from
the transformation property of $\tilde T_{00}$; Eqs.\
(\ref{EM-gauge-3}), (\ref{EM-gauge-4}) follow from $\tilde T_{0i}$;
Eqs.\ (\ref{EM-gauge-5}), (\ref{EM-gauge-6}) follow from $\tilde
T_{ij}$. The transformation functions $\varrho^{(2)}$ and
$v^{(2)}_i$ are not determined, see below.

If we take the spatial $C$-gauge, i.e., set $\xi^{(2)}_i \equiv 0$,
we have \bea
   & & \hat \varrho = \varrho
       + {1 \over c^2} \varrho^{(2)}, \quad
       \hat \Pi = \Pi
       - {\varrho^{(2)} \over \varrho}, \quad
       \hat v_i = v_i
       + {1 \over c^2} v^{(2)}_i, \quad
       \hat Q_i = Q_i - \varrho v^{(2)}_i, \quad
       \hat p = p + {\cal O}^{-4}, \quad
       \hat \Pi_{ij} = \Pi_{ij} + {\cal O}^{-4}.
   \label{EM-gauge2-6}
\eea Thus, we have \bea
   & & \hat \varrho \left( 1 + {1 \over c^2} \hat \Pi \right)
       = \varrho \left( 1 + {1 \over c^2} \Pi \right), \quad
       \hat v_i + {1 \over c^2} {1 \over \varrho} \hat Q_i
       = v_i + {1 \over c^2} {1 \over \varrho} Q_i.
\eea Although the gauge transformation properties of $\varrho$ and
$\Pi$ are not determined individually, the energy density $\mu
\equiv \varrho \left( 1 + \Pi/c^2 \right)$ is gauge-invariant under
our $C$-gauge. {}For vanishing $\Pi$ we have $\varrho^{(2)} = 0$.
Similarly, the gauge transformation properties of $v_i$ and $Q_i$
are not determined individually. {}For vanishing flux term in any
coordinate, $Q_i = 0$, we have $v^{(2)}_i = 0$, thus $\hat v_i = v_i
+ {\cal O}^{-4}$. Thus, the gauge transformation property of $v_i$
is determined only for vanishing flux term. {}From Eq.\
(\ref{varrho*-def}) we have \bea
   & & \hat \varrho^*
       = \varrho^* + {1 \over c^2} \varrho^{(2)}, \quad
       \hat v^*_i
       = v^*_i
       + {1 \over c^2} \left( {1 \over a}
       \xi^{(4)0}_{\;\;\;\;\;\;,i}
       - {\varrho^{(2)} \over \varrho} v_i \right).
\eea

We can check that up to the 1PN order Einstein's equations and the
energy and momentum conservation equations in Sec.\
\ref{sec:Equations} are invariant under the gauge transformation. In
the following we introduce several temporal gauge conditions each of
which fixes the temporal gauge mode completely. In relativistic
gravity, gauge conditions consist of four non-tensorial relations
imposed on the metric tensor or the energy-momentum tensor. Our
purpose is to employ the temporal gauge (slicing) condition to make
the resulting equations simple for mathematical/numerical treatment.
Depending on the situation we can also take alternative spatial
gauge conditions for that purpose. In the following we {\it impose}
the spatial $C$-gauge in Eq.\ (\ref{C-gauge}), and set $K = 0$.

\subsection{Chandrasekhar's gauge}
                                                 \label{sec:CHG}

Chandrasekhar's temporal gauge condition in his Eq.\ (24) of
\cite{Chandrasekhar-1965} corresponds to setting the 1PN part of
\bea
   & & \tilde g^{ij} \left( \tilde g_{0i,j}
       - {1 \over 2} \tilde g_{ij,0} \right)
       = - {1 \over c} 3 {\dot a \over a}
       - {1 \over c^3} \left( {1 \over a} P^i_{\;\;|i}
       + 3 \dot U \right)
       + L^{-1} {\cal O}^{-5},
\eea equal to zero; in the literature this is often called the
standard PN gauge \cite{Blanchet-etal-1990}. We take \bea
   & & {1 \over a} P^i_{\;\; |i} + 3 \dot U + m {\dot a \over a} U = 0,
   \label{CHG}
\eea as Chandrasekhar's gauge.  We used the freedom to add an
arbitrary $m (\dot a/a) U$ term with $m$, a real number. In this
case Eqs.\ (\ref{Pert3}), (\ref{Pert4}) give \bea
   {\Delta \over a^2} P_i
   &=& - 16 \pi G \varrho v_i
       + {1 \over a} \left[ \dot U
       - \left( m - 4 \right) {\dot a \over a} U \right]_{,i},
   \label{Pert3-CHG} \\
   {\Delta \over a^2} U
       + 4 \pi G \left( \varrho - \varrho_b \right)
   &+& {1 \over c^2} \Bigg\{
       2 {\Delta \over a^2} \Phi
       - \left( m - 3 \right) {\dot a \over a} \dot U
       + \left[ \left( 6 - m \right) {\ddot a \over a}
       - m {\dot a^2 \over a^2} \right] U
   \nonumber \\
   & &
       + 8 \pi G \left[ \varrho v^2
       + {1 \over 2} \left( \varrho \Pi - \varrho_b \Pi_b \right)
       + U \left( \varrho - \varrho_b \right)
       + {3 \over 2} \left( p - p_b \right) \right]
       \Bigg\} = 0.
   \label{Pert4-CHG}
\eea Therefore, $U$, $P_i$, and $\Phi$ are determined by Eqs.\
(\ref{Pert1}), (\ref{Pert3-CHG}), and (\ref{Pert4-CHG}). The
variable $\dot U$ can be determined from Eq.\ (\ref{Pert5}). This
completes our 1PN scheme based on Chandrasekhar's gauge. When we
handle this complete set of 1PN equations numerically, we should
monitor whether the chosen gauge condition is satisfied always; this
could be used to control the numerical accuracy.

\subsection{Uniform-expansion gauge}
                                                \label{sec:UEG}

The expansion scalar of the normal-frame vector, $\tilde \theta
\equiv \tilde n^c_{\;\;;c}$, is given in Eq.\ (\ref{theta-n-frame}).
It is the same as the trace of extrinsic curvature $\bar K$ with a
minus sign, see Eq.\ (\ref{ADM-extrinsic-curvature}). Taking the 1PN
part of $\bar K$ equal to zero \bea
   & & {1 \over a} P^i_{\;\; |i} + 3 \dot U + 3 {\dot a \over a} U
       \equiv 0,
   \label{UEG}
\eea can be naturally called the uniform-expansion gauge. In the
literature it is often called the ADM gauge
\cite{Blanchet-etal-1990}, or the maximal slicing condition in
numerical relativity \cite{Smarr-York-1978}. This condition
corresponds to the $m = 3$ case of Chandrasekhar's gauge in Eq.\
(\ref{CHG}).

\subsection{Transverse-shear gauge}
                                                \label{sec:TSG}

The shear of the normal-frame vector is given in Eq.\
(\ref{shear-n-frame}). Thus, the gauge condition \bea
   & & P^i_{\;\; |i} \equiv 0,
   \label{TSG}
\eea can be called the transverse-shear gauge. In this case Eqs.\
(\ref{Pert3}), (\ref{Pert4}) give \bea
   {\Delta \over a^2} P_i
   &=& - 16 \pi G \varrho v_i
       + {4 \over a} \left( \dot U + {\dot a \over a} U \right)_{,i},
   \label{Pert3-TSG} \\
   {\Delta \over a^2} U
       + 4 \pi G \left( \varrho - \varrho_b \right)
   &+& {1 \over c^2} \Bigg\{
       2 {\Delta \over a^2} \Phi
       + 3 \ddot U
       + 9 {\dot a \over a} \dot U
       + 6 {\ddot a \over a} U
   \nonumber \\
   & &
       + 8 \pi G \left[ \varrho v^2
       + {1 \over 2} \left( \varrho \Pi - \varrho_b \Pi_b \right)
       + U \left( \varrho - \varrho_b \right)
       + {3 \over 2} \left( p - p_b \right) \right]
       \Bigg\} = 0.
   \label{Pert4-TSG}
\eea Therefore, $U$, $P_i$, and $\Phi$ are determined by Eqs.\
(\ref{Pert1}), (\ref{Pert3-TSG}), and (\ref{Pert4-TSG}). $\dot U$
can be determined from Eq.\ (\ref{Pert5}). We still have $\ddot U$
which can be determined similarly by taking the time derivative of
Eq.\ (\ref{Pert5}) and by using the background and the Newtonian
equations in Eqs.\ (\ref{BG-eq2-c})-(\ref{Poisson-eq-c}).

\subsection{Harmonic gauge}
                                                \label{sec:HG}

We have \bea
   & & \tilde g^{ab} \tilde \Gamma^0_{ab}
       = - {1 \over \sqrt{- \tilde g}}
       \left( \sqrt{- \tilde g} \tilde g^{0a} \right)_{,a}
       = { 1\over c} 3 {\dot a \over a}
       + {1 \over c^3} \left(
       {1 \over a} P^i_{\;\; |i}
       + 4 \dot U
       + 6 {\dot a \over a} U
       \right)
       + L^{-1} {\cal O}^{-5}.
   \label{HG0}
\eea The well known harmonic gauge condition sets the 1PN part of
Eq.\ (\ref{HG0}) equal to zero, thus we take \bea
   & & {1 \over a} P^i_{\;\; |i} + 4 \dot U
       + m {\dot a \over a} U
       \equiv 0,
   \label{HG}
\eea where we used a freedom to add an arbitrary $m(\dot a/a) U$
term. Combined with Eq.\ (\ref{spatial-harmonic}) the full harmonic
gauge condition can be expressed by setting the 1PN parts of $\tilde
g^{ab} \tilde \Gamma^c_{ab}$ equal to zero. In the perturbation
approach the harmonic gauge (often called de Donder gauge) is known
to be a bad choice because the condition involves derivatives of the
metric variables; this leads to an incomplete gauge fixing and
consequently we would have to handle higher order differential
equations which is unnecessary, see the Appendix in
\cite{H-IF-1993}. In this gauge Eqs.\ (\ref{Pert3}), (\ref{Pert4})
give \bea
   {\Delta \over a^2} P_i
   &=& - 16 \pi G \varrho v_i
       - \left( m - 4 \right) {\dot a \over a} {1 \over a} U_{,i},
   \label{Pert3-HG} \\
   {\Delta \over a^2} U
       + 4 \pi G \left( \varrho - \varrho_b \right)
   &+& {1 \over c^2} \Bigg\{
       2 {\Delta \over a^2} \Phi
       - \ddot U
       - \left( m - 1 \right) {\dot a \over a} \dot U
       + \left[ \left( 6 - m \right) {\ddot a \over a}
       - m {\dot a^2 \over a^2} \right] U
   \nonumber \\
   & &
       + 8 \pi G \left[ \varrho v^2
       + {1 \over 2} \left( \varrho \Pi - \varrho_b \Pi_b \right)
       + U \left( \varrho - \varrho_b \right)
       + {3 \over 2} \left( p - p_b \right) \right]
       \Bigg\} = 0.
   \label{Pert4-HG}
\eea Therefore, $U$, $P_i$, and $\Phi$ are determined by Eqs.\
(\ref{Pert1}), (\ref{Pert3-HG}), and (\ref{Pert4-HG}). The variable
$\dot U$ can be determined from Eq.\ (\ref{Pert5}). We also have a
$\ddot U$ term in Eq.\ (\ref{Pert4-HG}) which might demand for a
more involved numerical implementation as explained below Eq.\
(\ref{Pert4-TSG}). However, its presence makes the propagating
nature of the 1PN order metric fluctuations apparent in this gauge
condition. This time-delayed propagation of the gravitational field
in the 1PN approximation can be compared with the
action-at-a-distance nature of Poisson's equation in the Newtonian
order in Eq.\ (\ref{Pert4-HG}). We anticipate that the relativistic
time-delayed propagation could lead to a secular (time cumulative)
effect. This could be important even in the case in which the 1PN
correction terms are small compared with the Newtonian terms; in the
large-scale clustering regions we would expect $(v/c)^2 \sim
GM/(Rc^2) \simeq 10^{-6} \sim 10^{-4}$, see Sec.
\ref{sec:Discussion}. Using $\tilde g_{00} \equiv -1 + 2 {\bf U} /
c^2$ we have \bea
   & & {\bf U} \equiv U
       + {1 \over c^2} \left( - U^2 + 2 \Phi \right) + c^2 {\cal O}^{-4},
\eea thus \bea
       \Box {\bf U}
       \equiv {\bf U}^{;c}_{\;\;\; c}
   &=& {\Delta \over a^2} {\bf U}
       - {1 \over c^2} \left(
       \ddot {\bf U} + 3 {\dot a \over a} \dot {\bf U}
       + 2 {\bf U} {\Delta \over a^2} {\bf U} \right)
       + {\cal T}^{-2} {\cal O}^{-4}
   \nonumber \\
   &=&
       {\Delta \over a^2} U
       - {1 \over c^2} \left[ {\Delta \over a^2} \left( U^2 - 2 \Phi \right)
       + \ddot U + 3 {\dot a \over a} \dot U
       + 2 U {\Delta \over a^2} U \right]
       + {\cal T}^{-2} {\cal O}^{-4},
   \label{Box}
\eea and Eq.\ (\ref{Pert4-HG}) can be written as \bea
   & & \Box {\bf U}
       + 4 \pi G \left( \varrho - \varrho_b \right)
   \nonumber \\
   & & \qquad
       + {1 \over c^2} \left\{
       {\Delta \over a^2} {\bf U}^2
       - \left( m - 4 \right) {\dot a \over a} \dot {\bf U}
       + \left[ \left( 6 - m \right) {\ddot a \over a}
       - m {\dot a^2 \over a^2} \right] {\bf U}
       + 8 \pi G \left[ \varrho v^2
       + {1 \over 2} \left( \varrho \Pi - \varrho_b \Pi_b \right)
       + {3 \over 2} \left( p - p_b \right) \right]
       \right\} = 0.
   \label{Pert4-HG2}
\eea

Thus, in the harmonic gauge condition the propagation speed of the
gravitational potential ${\bf U}$ is the same as the speed of light
$c$. (When we mention the propagation speed, we are considering the
D'Alembertian part of the wave equation ignoring the nonlinear terms
and background expansion. In this case we have $\Box {\bf U} =
a^{-2} \Delta {\bf U} - c^{-2} \ddot {\bf U}$, and for $K = 0$,
$\Delta$ becomes the ordinary Laplacian in flat space.) Apparently,
the wave speed of the metric (potential) can take an arbitrary value
depending on the temporal gauge condition we choose. {}For example,
if we take \bea
   & & {1 \over a} P^i_{\;\; |i} + n \dot U
       + m {\dot a \over a} U
       \equiv 0,
   \label{GG}
\eea as the gauge condition, with $n$ and $m$ real numbers, Eqs.\
(\ref{Pert3}), (\ref{Pert4}) give \bea
   {\Delta \over a^2} P_i
   &=& - 16 \pi G \varrho v_i
       - {1 \over a} \left[ \left( n - 4 \right) \dot U
       + \left( m - 4 \right) {\dot a \over a} U \right]_{,i},
   \label{Pert3-GG} \\
   {\Delta \over a^2} U
       + 4 \pi G \left( \varrho - \varrho_b \right)
   &+& {1 \over c^2} \Bigg\{
       2 {\Delta \over a^2} \Phi
       - \left( n - 3 \right) \ddot U
       - \left( 2n + m -9 \right) {\dot a \over a} \dot U
       + \left[ \left( 6 - m \right) {\ddot a \over a}
       - m {\dot a^2 \over a^2} \right] U
   \nonumber \\
   & &
       + 8 \pi G \left[ \varrho v^2
       + {1 \over 2} \left( \varrho \Pi - \varrho_b \Pi_b \right)
       + U \left( \varrho - \varrho_b \right)
       + {3 \over 2} \left( p - p_b \right) \right]
       \Bigg\} = 0.
   \label{Pert4-GG}
\eea In this case, for $n \ge 3$, the speed of propagation
corresponds to \bea
   & & {c \over \sqrt{n - 3}},
   \label{propagation-speed}
\eea which can take an {\it arbitrary} value depending on our choice
of the value of $n$. It becomes $c$ for $n = 4$ (e.g., the harmonic
gauge), and infinity for $n = 3$ (e.g., Chandrasekhar's gauge and
the uniform-expansion gauge). In the case of the transverse-shear
gauge we have $n = 0$, thus Eq.\ (\ref{Pert4-TSG}) is no longer a
wave equation.

\subsection{Transformation between two gauges}
                                                   \label{sec:GT}

Now, let us show how we can relate the equations and solutions known
in one gauge condition to the ones in any other gauge condition. Our
spatial $C$-gauge condition already fixed the gauge transformation
function $\xi^{(2)}_i \equiv 0$, and we have $\xi^{(2)0} \equiv 0$.
Under the remaining (temporal) gauge transformation \bea
   & & \hat t = t + {1 \over c^4} \xi^{(4)0}, \quad
       \hat x^i = x^i,
\eea we have \bea
   & & \hat \Phi = \Phi + {1 \over 2} \dot \xi^{(4)0}, \quad
       \hat P_i = P_i - {1 \over a} \xi^{(4)0}_{\;\;\;\;\;\;,i},
   \label{GT-2}
\eea and all the other variables are gauge-invariant. As an example,
let us consider the two gauge conditions used in Sec.\ \ref{sec:TSG}
and Sec.\ \ref{sec:HG}. Let us assume that the $x^a$ coordinate is
the transverse-shear gauge in Sec.\ \ref{sec:TSG}, and $\hat x^a$ is
the harmonic gauge in Sec.\ \ref{sec:HG}. Thus, we have $P^i_{\;\;
|i} \equiv 0$ in $x^a$, and $4 \dot U + m {\dot a \over a} U
       + a^{-1} P^i_{\;\; |i} \equiv 0$ in $\hat x^a$.
{}From Eq.\ (\ref{GT-2}) we have \bea
   & & - \left( 4 \dot U
       + m {\dot a \over a} U \right)
       = 0 - {\Delta \over a^2} \xi^{(4)0},
\eea thus \bea
   & & {\Delta \over a^2} \xi^{(4)0}
       = 4 \dot U
       + m {\dot a \over a} U.
   \label{GT-3}
\eea Thus, when we transform the result known in the
transverse-shear gauge to the harmonic gauge, we use Eq.\
(\ref{GT-2}) with $\xi^{(4)0}$ given in Eq.\ (\ref{GT-3}). All the
other variables are invariant under the gauge transformation. We can
show that under this transformation Eqs.\ (\ref{Pert3-TSG}),
(\ref{Pert4-TSG}) become Eqs.\ (\ref{Pert3-HG}), (\ref{Pert4-HG}),
respectively. In this way, if we know the solution in any gauge
condition the rest of the solutions in other gauge conditions can be
simply derived without solving the equations again. Thus, as in
other gauge theories (like Maxwell's or Yang-Mills' theories) the
gauge condition should be deployed to our advantage in handling the
problem mathematically.

%
%

%
%
\section{Discussion}
                                          \label{sec:Discussion}

Here we compare the PN approach with the perturbative one. The
perturbation analysis is based on  the perturbation expansion of the
metric and energy-momentum variables in a given background. All
perturbation variables are assumed to be small. In linear order
perturbations we keep only the first-order deviations from the
background \cite{Lifshitz-1946,Linear-perturbations,Bardeen-1980},
whereas in the weakly nonlinear perturbations we keep higher-order
deviations to the desired order \cite{NL}. Contrary to the PN
approach the perturbation analysis is applicable in the strong
gravity regime and on all cosmological scales as long as the
perturbations are linear or weakly nonlinear.

Meanwhile, in the PN approach, by assuming weak gravitational fields
and slow motions, we try to provide the general relativistic
correction terms for the Newtonian equations of motion. Thus, in the
PN approach, in fact, we abandon the geometric spirit of general
relativity and recover the concept of absolute space and absolute
time. Although this could be regarded as a shortcoming of the PN
approach, in this way it provides the relativistic effects in forms
of the correction terms to the well known Newtonian equations, thus
enabling us to use simpler conventional (numerical) treatment. We
expand the metric and the energy-momentum variables in powers of
$v/c$ in a given background spacetime. In nearly virialized system
we have $GM / (R c^2) \sim (v / c)^2$ which is assumed to be small.
Thus, no strong gravity situation is allowed and the results are
valid inside horizon $\sqrt{GM / (R c^2)} \sim {R / (c/H)} < 1$. In
comparison to the perturbation approach, however, the resulting
equations in the 1PN approximation can be regarded as fully
nonlinear. As in the perturbative case, even in our cosmological PN
approach we {\it assume} the presence of a Robertson-Walker
cosmological background.

As we have summarized in the introduction, our studies of the weakly
nonlinear regime of a zero-pressure cosmological medium showed that
the Newtonian equations are quite successful even near the horizon
scale where the fluctuations are supposed to be near linear stage.
We have shown that to the second order in perturbations, except for
the presence of gravitational waves, the relativistic equations
coincide exactly with the Newtonian ones. The pure relativistic
correction terms appearing in the third order are independent of the
presence of the horizon and are small compared with the second-order
terms by a factor $\delta T/T \sim 10^{-5}$, thus negligible.

In order to properly estimate the relativistic effects in the
evolution of large-scale cosmic structures we have to implement our
equations in a hydrodynamic cosmological numerical simulation. The
PN correction terms are $(v/c)^2$ or $GM/(Rc^2)$ orders smaller than
the Newtonian terms. In Newtonian numerical simulations the maximum
large scale velocity field of a cluster flow reaches nearly
$3000km/sec$ and the typical value for the velocity is about an
order of magnitude smaller than this \cite{Kang-etal-1996}. Thus, we
may estimate the 1PN effect to be of the order $(v/c)^2 \simeq
10^{-6} \sim 10^{-4}$, thus quite small. We already mentioned that
the PN approximation is applicable inside the horizon only.
Considering the action-at-a-distance nature of the Newtonian gravity
theory it is important to check the domain of validity of the
Newtonian theory in the nonlinear evolution of cosmological
structures. The PN approach provides a way to find out the
relativistic effects in such a regime. We anticipate that the
propagating nature of the gravitational field with finite speed in
relativistic gravity theory, compared with the instantaneous
propagation in Newton's theory, could lead to accumulative (or
secular) relativistic effects. In order to estimate relativistic
effects it would be appropriate to consider a single cold dark
matter component as a zero-pressure fluid without internal energy.
In such a case our equations in Sec.\ \ref{sec:IF} with $p = 0 =
\Pi$ and the metric variables $U$, $\Phi$, and $P_i$ presented in
various gauge conditions in Sec.\ \ref{sec:CHG}-\ref{sec:HG} would
provide a complete set of equations expressed in various forms.



Large-scale structures are still near the linear regime, and due to
the enhanced amplitude of the initial mass power spectrum in the
small scale the gravitational evolution causes nonlinear regions to
begin in the small scale and to propagate to larger scales
\cite{Smith-etal-2003}. The current belief is that in small scale
structures, where the nonlinearity is important, the dynamical
time-scale is much longer than the light travel time over this
scale, thus the time dilation effect from relativistic gravity is
not important. The 1PN order relativistic effects could be important
in the tidal interactions among clusters of galaxies, where the
dynamical time-scale could become substantial compared with the
light crossing time of the scales involved \cite{Ryu-2005}.

We can perform numerical simulations based on any two different
gauge conditions. The results should coincide after making a gauge
transformation between the two gauges as in Sec.\ \ref{sec:GT}.
Also, the two simulations should give identical result for any given
gauge-invariant variable. These might provide a way to check the
numerical accuracy of the simulations. Analyses based on different
gauge conditions would lead to the different results; this is in the
sense that a given variable evaluated in two different gauges are
actually two different variables, unless the original variable is
gauge-invariant. Even for the temporal gauge condition (after fixing
the spatial $C$-gauge) there are infinitely many different gauge
conditions available. Since each of the gauge conditions displayed
in Sec.\ \ref{sec:CHG}-\ref{sec:HG} fixes the temporal gauge mode
completely (and the spatial gauge modes are already completely fixed
by our spatial $C$-gauge), all remaining variables are equivalently
gauge-invariant. Thus, apparently, the gauge-invariance does not
guarantee us to associate the variables with physically measurable
quantities. The identification of physically measurable quantities
out of infinitely many gauge-invariant candidates is still an open
issue which remains to be addressed. The gauge invariance assures
that the value should not depend on the gauge we take and this fact
can be used to check and control the numerical accuracy of the
simulations.

In a related context, we have shown that the propagation speed of
the gravitational potential depends on the temporal gauge condition
we take, see Eq.\ (\ref{propagation-speed}). A similar situation
occurs in classical electrodynamics. The propagation speeds of the
electromagnetic scalar and vector potentials are the speed of light
$c$ in the Lorenz gauge, whereas that of scalar potential becomes
infinite in the Coulomb gauge. In electromagnetism the issue is well
resolved by showing that the electric and magnetic fields propagate
with $c$ independently of the adopted gauge condition
\cite{Jackson-2002}. In our 1PN approach, however, we have {\it not}
been able to resolve the case. In the PN case this is related to
identifying the physically relevant gauge condition out of
infinitely many different gauge conditions available (which are all
gauge-invariant), an issue we have described in the previous
paragraph. One difference compared to electromagnetism is that the
PN approach addresses fully nonlinear gravitational dynamics whereas
electromagnetics concerns linear processes. We wish to address these
important issues in a future occasion.

Just like the solar system tests of Einstein's gravity theory, the
nonlinear evolution of the large-scale cosmic structure could
provide another regime where the gravitational field is weak and the
motions are slow so that the post-Newtonian approximation would be
practically adequate to describe the ever-present relativistic
effects using Newtonian-like equations. The problem is whether such
effects are significant enough to be detected in future observations
and numerical experiments. This important issue is left for future
studies. We hope that our set of 1PN order equations and our
strategy for using those equations would be useful for such studies
in the cosmological context. In more realistic cosmological
situations we have dust and cold dark matter which can be
approximated by two zero-pressure ideal fluids. We can also include
pressure and dissipation effects in the case of dust. The
multi-component situation of cosmological 1PN hydrodynamics will be
considered in later extensions of this work. Extending our
formulation to higher order PN approximation is also an apparent
next step which would be tedious but straightforward.

%
%
\section*{Acknowledgments}

We thank Prof. Dongsu Ryu and Dr. Juhan Kim for useful discussions
from the perspective of possible numerical implementations. DP
wishes to thank Prof. M. Pohl for his support of this project. HN
was supported by grants No. R04-2003-10004-0 from the Basic Research
Program of the Korea Science and Engineering Foundation. JH was
supported by the Korea Research Foundation Grant No.
2003-015-C00253.

%
%

\end{document}